\documentclass[journal,draftcls,onecolumn,12pt,twoside]{IEEEtranTCOM}
\usepackage{graphicx}
\usepackage{amssymb}
\usepackage{amsmath}
\usepackage[table]{xcolor}
\DeclareMathOperator{\Tr}{Tr}

\DeclareMathOperator*{\argmax}{arg\,max}
\DeclareMathOperator*{\argmin}{arg\,min}
\usepackage{algorithmic}
\usepackage{array}
\usepackage[caption=false,font=footnotesize]{subfig}
\usepackage{stfloats}
\usepackage{url}
\usepackage{caption}
\usepackage[short]{optidef}
\usepackage{bm}
\usepackage{makecell}
\usepackage{algorithmic}
\usepackage{booktabs}
\usepackage{adjustbox}
\usepackage{comment}
\usepackage{multirow}
\usepackage{cite}
\usepackage[linesnumbered,ruled,vlined]{algorithm2e}
\usepackage{setspace}
\def\BibTeX{{\rm B\kern-.05em{\sc i\kern-.025em b}\kern-.08em
    T\kern-.1667em\lower.7ex\hbox{E}\kern-.125emX}}

\SetKwInput{KwInput}{Input}                
\SetKwInput{KwOutput}{Output}              
\ifCLASSINFOpdf
\else
\fi

\begin{document}
\title{Model-based Deep Learning Receiver Design for Rate-Splitting Multiple Access}

\author{Rafael~Cerna~Loli,~\IEEEmembership{Student~Member,~IEEE},
        Onur~Dizdar,~\IEEEmembership{Member,~IEEE},
        Bruno~Clerckx,~\IEEEmembership{Fellow,~IEEE},
        and~Cong~Ling,~\IEEEmembership{Member,~IEEE}
\thanks{R. Cerna Loli is supported by a grant provided by the Defence Science and Technology Laboratory (Dstl) Communications and Networks Research Programme.}
\thanks{
R. Cerna Loli, O. Dizdar, and C. Ling are with  the Department of Electrical and Electronic Engineering at Imperial College London, London SW7
2AZ, UK. B. Clerckx is with the Department of Electrical and Electronic Engineering at Imperial College London, London SW7 2AZ, UK and with Silicon Austria Labs (SAL), Graz A-8010, Austria. (email: rafael.cerna-loli19@imperial.ac.uk;  o.dizdar@imperial.ac.uk; b.clerckx@imperial.ac.uk; c.ling@imperial.ac.uk).}}

\maketitle

\begin{abstract}
Effective and adaptive interference management is required in next generation wireless communication systems. To address this challenge, Rate-Splitting Multiple Access (RSMA), relying on multi-antenna rate-splitting (RS) at the transmitter and successive interference cancellation (SIC) at the receivers, has been intensively studied in recent years, albeit mostly under the assumption of perfect Channel State Information at the Receiver (CSIR) and ideal capacity-achieving modulation and coding schemes. To assess its practical performance, benefits, and limits under more realistic conditions, this work proposes a novel design for a practical RSMA receiver based on model-based deep learning (MBDL) methods, which aims to unite the simple structure of the conventional SIC receiver and the robustness and model agnosticism of deep learning techniques. The MBDL receiver is evaluated in terms of uncoded Symbol Error Rate (SER), throughput performance  through Link-Level Simulations (LLS), and average training overhead. Also, a comparison with the SIC receiver, with perfect and imperfect CSIR, is given. Results reveal that the MBDL receiver outperforms by a significant margin the SIC receiver with imperfect CSIR, due to its ability to generate on demand non-linear symbol detection boundaries in a pure data-driven manner.  
\end{abstract}

\begin{IEEEkeywords}
Rate-splitting multiple access (RSMA), robust interference management, imperfect channel state information (CSI) at the receiver (CSIR), model-based deep-learning (MBDL), deep neural network (DNN).
\end{IEEEkeywords}

\IEEEpeerreviewmaketitle

\section{Introduction}

Multiple access schemes and multi-user multiple-input multiple-output (MU-MIMO) antenna processing for efficient multi-user interference management have become hotspot research areas to fulfil the requirements of next generation wireless communications technologies \cite{rs_overview}. However, research efforts often consider that perfect Channel State Information at the Transmitter (CSIT) and perfect Channel State Information at the Transmitter (CSIR) are available. Since channel estimation errors are inevitable in practice, this assumption leads to sub-optimal adaptations of the developed techniques under perfect CSIT and perfect CSIR into scenarios with imperfect CSIT and imperfect CSIR \cite{joudeh,lina_dpc}.

\begin{figure*}[t]
		\centering
		\includegraphics[width=\textwidth]{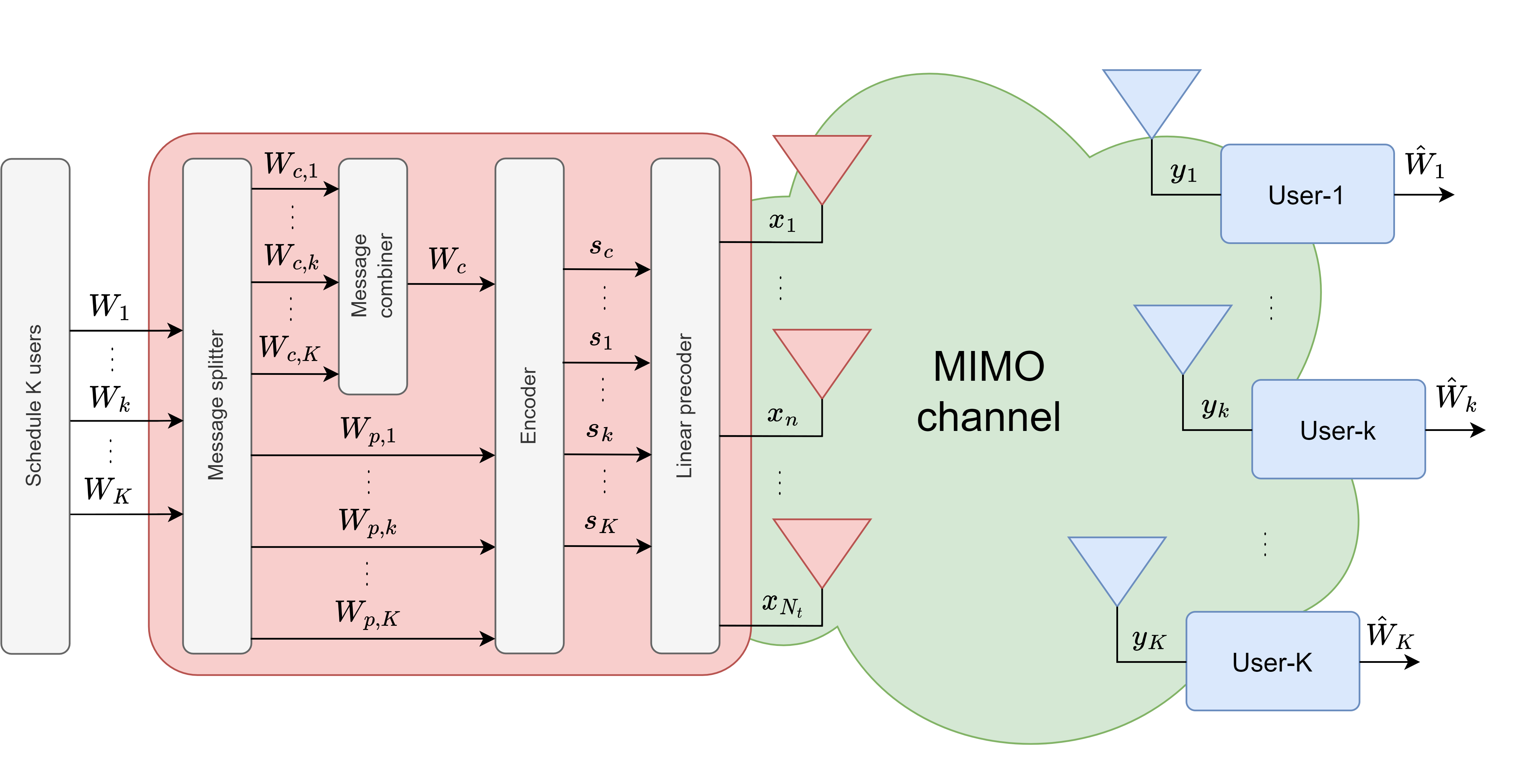}
		\caption{1-Layer RSMA system model \cite{rs_overview}.}
		\label{fig:rsma_system_model}
\end{figure*}

In recent years, a new multi-user transmission approach has been introduced, Rate-Splitting Multiple Access (RSMA), which relies on linearly or non-linearly precoded Rate-Splitting (RS) to partially decode the multi-user interference (MUI) and partially treat it as noise \cite{eurasip}. As depicted in Fig. \ref{fig:rsma_system_model}, such a processing is achieved by first splitting the user messages and encoding them into common and private streams. Then, all users decode the common stream and, after employing Successive Interference Cancellation (SIC), each of them decodes its intended private stream while considering the private streams of the other users as noise. Owing to its generalized structure, RSMA unifies other apparently non-related transmission schemes, such as Space Division Multiple Access (SDMA), which fully treats MUI as noise, Non-Orthogonal Multiple Access (NOMA), which fully decodes MUI, Orthogonal Multiple Access (OMA), which avoids MUI by transmitting in different radio resources, and physical-layer multicasting. From an information-theoretic perspective, this strategy translates into an increase in the Degrees of Freedom (DoFs) for each user and an increase in the total system Sum-Rate (SR) \cite{joudeh}. It has also been demonstrated that RSMA offers significant benefits in terms of spectral efficiency, reliability, ability to comply with Quality of Service (QoS) constraints, and robustness against CSIT errors \cite{rs_6g}. 
Even though there are numerous studies on RSMA considering both perfect CSIT \cite{eurasip,perfect_1,perfect_3, perfect_4,perfect_5} and imperfect CSIT \cite{rs_overview,joudeh,lina_dpc,joudeh_2,imperfect_1,imperfect_2,imperfect_3,imperfect_4,imperfect_5,imperfect_6}, interference management analysis and algorithms to maximize the total sum-rate have mostly been studied from the transmitter side, assuming that the receiver possesses perfect CSIR and is thus able to perform perfect SIC. Although initial attempts to depart from these idealistic conditions were made in \cite{onur_ref}, \cite{jihye}, there is still an urgent need to design robust receivers to operate under the practical constraints of imperfect CSIR, finite constellation modulation schemes, and finite-length channel codes. In this section, we first review the limitations of current RSMA receiver architectures. Then, we discuss how deep learning (DL) methods can enhance the performance of wireless communication systems and also present the three specific challenges that DL-based wireless receivers face in a practical implementation. Finally, we describe the motivation to design an optimized model-based deep learning wireless RSMA receiver and present the main contributions of this work at the end of the section. The main abbreviations employed in this paper are detailed in Table \ref{acronym_table}.
    \begin{table}[h]
    \centering
    \caption{List of abbreviations.}
    \label{acronym_table}
    \resizebox{\columnwidth}{!}{
    \begin{tabular}{|l|l|}
    \hline
     AMC & Adaptive Modulation and Coding \\
     BPSK & Binary Phase Shift Keying \\
     CSCG & Circularly Symmetric Complex Gaussian \\
     CSI & Channel State Information \\
     CSIR & Channel State Information at the Receiver \\
     CSIT & Channel State Information at the Transmitter\\
     DL & Deep Learning \\ 
     DNN & Deep Neural Network \\
     DoF & Degrees of Freedom \\
     FC & Fully Connected \\
     ISIC & Iterative Soft Interference Cancellation \\
     LLS & Link-Level Simulations \\
     LPR & Log-Probability Ratio \\
     MAP & Maximum-a-posteriori probability \\
     MBDL & Model-Based Deep Learning \\
     \hline
    \end{tabular}
    \begin{tabular}{|l|l|}
    \hline
     ML & Maximum Likelihood \\
     MUI & Multi-User Interference \\
     NOMA & Non-Orthogonal Multiple Access \\
     OMA & Orthogonal Multiple Access \\
     PHY-Layer & Physical Layer \\
     QAM & Quadrature Amplitude Modulation \\
     QoS & Quality of Service \\
     QPSK & Quadrature Phase Shift Keying \\
     RS & Rate-Splitting \\
     RSMA & Rate-Splitting Multiple Access \\
     SCL & Successive Cancellation List \\
     SDMA & Space Division Multiple Access \\
     SER & Symbol Error Rate \\
     SINR & Signal-to-Interference-and-Noise Ratio \\
     SIC & Successive Interference Cancellation \\
    \hline
    \end{tabular}
    }
    \end{table}

\subsection{Limitations of current RSMA receiver architectures}
It has been mentioned in \cite{comsoc_wtc,onur_ref_2} that RSMA is not limited to the use of SIC receivers, but rather it is also suited for other types of architectures, such as joint decoding and turbo decoding receivers. However, to the best of our knowledge, only \cite{jd_1, jd_2} analyze the performance of an alternate scheme with joint decoding for RSMA, albeit from a purely information-theoretic transmitter-side perspective and without proposing practical and suitable channel codes and symbol detection methods. Additionally, precoder optimization is done assuming perfect CSIT and perfect CSIR, and later adapting it into scenarios with imperfect CSIT and perfect CSIR. This renders the optimization algorithm sub-optimal compared to designs considering the latter assumptions \cite{lina_dpc}.

In order to understand the limitations of conventional receiver design, it is necessary to evaluate the problem from a more general perspective that encompasses RSMA and other multi-user multiple-access schemes. Based on that, it is evident that conventional receiver architectures depend on predefined models and specific assumptions about the channel and nature of the interference and that, in order to keep these models tractable, it is necessary to make assumptions to describe the signal propagation, noise, interference type, etc \cite{model-based-dl}. Specifically, regarding the SIC receiver, the challenges in a practical deployment are twofold \cite{lmmse}: First, most past works about RSMA assume perfect CSIR to perform perfect interference cancellation, which is clearly not a realistic assumption and would result in critical decoding errors. Thus, a practical SIC receiver would still need to be designed and optimized under the consideration of error propagation due to imperfect interference cancellation. Second, the delay of decoding each stream one by one needs to be considered, although in 1-layer RSMA this may not be critical as each user only applies one SIC layer compared to the multiple SIC layers needed in NOMA receivers.  

Even though a joint decoding receiver would tackle the limitations of successive decoding with SIC layers, it would do so at the expense of increasing the receiver complexity. As will be explained in the next section, the optimum joint decoding receiver, the maximum a-posteriori probability (MAP) receiver, minimizes the decoded bit error rate but suffers from exponentially high complexity as the number of streams and their modulation order increase, which prohibits its use in a practical implementation \cite{map_1}. For this reason, the family of iterative decoding receivers have been proposed, such as the iterative soft interference cancellation (ISIC) receiver \cite{map_rule}, the iterative LMMSE detection receiver \cite{lmmse}, and the iterative message passing detection receiver \cite{message_passing}. These receivers, assuming perfect CSIR, perform joint decoding with near-MAP performance and lower complexity by performing multi-user detection (with or without interference cancellation) and refining the decoder output bit probability estimates iteratively \cite{deep_sic}. Compared to the SIC receiver, the iterative decoding receivers recover the user message with lower delay due to parallel decoding of the streams at the expense of a larger computational burden due to employing multi-core processing, and complexity as several iterations are required for convergence \cite{pic_2}. In the context of RSMA communications, the achievable latency decrease would need to be evaluated to determine if the complexity increase of iterative decoding receivers can be afforded, although this is not likely in 1-layer RSMA communications as only two streams need to be decoded at the receivers regardless of the total number of users in the system.

\subsection{Challenges of applying deep-learning to wireless communications}
Conventional receiver architectures are based on algorithms designed for specific mathematical channel models, which may be of statistical or deterministic nature, that aim to define the relationship between the transmitted and received signals. Consequently, estimation of the channel model parameters is necessary as these \textit{model-based algorithms} rely strongly on accurate prior \textit{model knowledge} and perform poorly if it is not accurately acquired \cite{model-based-dl}.

In contrast, DL has emerged as a promising, powerful and purely data-driven tool for wireless communications \cite{dl_comms_survey}. A major advantage of receivers based on DL methods is that they are able to directly extract meaningful information from the unknown channel solely on observations. Therefore, DL is naturally suited for scenarios in which the underlying mathematical channel model is unknown, its parameters cannot be acquired with precision, or when it is too complex to be characterized by model-based algorithms with low computational resources \cite{deep-learning-comms, dl_challenges}. However, the application of conventional DL techniques, where a single Deep Neural Network (DNN) is employed, in wireless receivers faces three significant challenges \cite{model-based-dl-2}:
\begin{enumerate}
    \item Though the main advantage of a DNN is its model-agnosticism, this also results in a generalized and complex network with numerous nodes and layers, which requires a large training set to learn a specific mapping. This, in turn, derives into a large computational burden at the receiver during the training phase.
    \item Transmitting a large training set would cause a large training overhead and delay the transmission of actual useful data to the communication users.
    \item Due to the time-varying and dynamic nature of wireless channels, periodical training of the DNNs at the receivers is required to account for any channel variations, which would make the transmission of large training sets highly impractical.
\end{enumerate}

\subsection{Motivation for an optimized model-based deep learning receiver design}
To address these three issues, \textit{model-based deep learning} (MBDL) has been introduced as an effort to marry the simplicity of conventional model-based algorithms with the model-agnosticism of DNNs \cite{model-based-dl,model-based-dl-2}. To achieve this, MBDL systems are implemented by replacing specific steps and computations in model-based algorithms that require accurate channel model knowledge with compact DNNs that require smaller training data sets compared to traditional DL systems with only one complex DNN. In the context of wireless receivers, previous MBDL adaptations of the SIC receiver have been proposed for uplink and downlink NOMA in \cite{noma_dl_1} and \cite{noma_dl_2}. In these works, specialized DNNs are used to perform interference cancellation and/or symbol classification, and it is demonstrated that this approach outperforms the conventional model-based SIC receiver for NOMA. However, their limitations are twofold:
\begin{enumerate}
    \item Although the DNNs used in the MBDL NOMA receivers are more compact compared to using a single DNN in the traditional DL manner, the provided simulation results are obtained by only contemplating lower-order modulation schemes (e.g. BPSK, QPSK, 16QAM) as the complexity of the DNNs and the required training set size still largely increase with the number of symbols in the constellation used to modulate the data streams. 
    \item For effective simultaneous training of MBDL NOMA receivers, the transmitted training data set must be generated using all possible symbol combinations among the individual user streams. Therefore, the training set size also increases exponentially with the total number of users.
\end{enumerate}

Thus, besides addressing the three challenges of applying DL methods to wireless receivers, the objective of the present work is to design an optimized MBDL receiver for RSMA that can also overcome the limitations of past MBDL receivers with the ability to serve multiple users simultaneously while employing high order modulation schemes.

\subsection{Contributions}
In this work, a novel design for an MBDL RSMA receiver is proposed in order to improve the performance of practical RSMA-based communications with imperfect CSIR. The main contributions are then summarized as follows:

\textit{First, }to the best of the knowledge of the authors, this is the \text{first paper} to propose a practical RSMA receiver design based on DL techniques. This new architecture incorporates MBDL techniques to take advantage of the ability of DNNs to learn the optimum channel mapping function in a purely data-driven approach (independent of any CSIR quality), in order to surpass the performance of the classical SIC receiver with imperfect CSIR. 

\textit{Second, }using the MBDL approach, only the computations in the SIC receiver that require CSIR knowledge are replaced with compact DNN banks to perform symbol detection of QAM symbols. Each bank is composed of two dedicated DNNs that take advantage of the symmetric nature of QAM constellations to estimate, in a two dimensional lookup table manner, in which row and column of the constellation the target symbol is located. In this way, the large classification problem of detecting symbols of high-order QAM schemes is divided into smaller ones to achieve a DNN complexity reduction compared to employing only a single DNN.

\textit{Third, }it is demonstrated that, by employing DNN banks to handle the row and column detection problems for each stream, a reduction in the total number of required training symbols can be achieved compared to employing single DNNs. Specifically, for QAM symbols that carry $M$ bits each, the single DNN requires enough training data to learn $2^M$ classes, whereas each DNN in the proposed architecture requires only enough training data to learn $2^{\frac{M}{2}}$ classes.

\textit{Fourth, }two training symbol patterns are introduced for the MBDL receiver as an effort to minimize the system training overhead. The first takes advantage of the unique common stream of RSMA communications. As each user only needs to know the training symbols of the common stream and its own private stream, generating a training signal composed of all possible symbol combinations between the common stream and the private stream of the user with the highest modulation order, and random symbol combinations for the rest of the private streams allows for a training set size reduction compared to transmitting all possible symbol combinations among streams. The second training symbol pattern takes advantage of the symmetry of QAM constellations. Under this scheme, only a small number of training symbols are generated, which are then interpolated at the receivers to complete the training set. In this way, the system training overhead is further minimized.

\textit{Fifth, }by numerical results, we demonstrate that the MBDL receiver can vastly outperform, in terms of uncoded Symbol Error Rate (SER), and throughput through coded transmissions, the SIC receiver with imperfect CSIR ans also approach the performance of the SIC receiver with perfect CSIR, even in an overloaded scenario when the number of users is larger than the number of transmit antennas.

\textit{Notation:} Scalars, vectors and matrices are denoted by standard, bold lower and upper case letters, respectively. $\mathbb{R}$ and $\mathbb{C}$ denote the real and complex domains. The transpose and Hermitian transpose operators are represented by $(.)^T$ and $(.)^H$, respectively. The expectation of a random variable is given by $\mathbb{E}\{.\}$. $\Re\{.\}$ and $\Im\{.\}$ indicate the real and imaginary parts of a complex number, $||.||_2$ is the $l_2$-norm operator, and $\max(.,.)$ is the operator that chooses the maximum between its input parameters.  Finally, $\mathtt{\sim}$ denotes "distributed as" and $\mathcal{CN}(0,\sigma^2)$ denotes
the Circularly Symmetric Complex Gaussian (CSCG) distribution with zero mean and variance $\sigma^2$.

\section{RSMA System Model and Receiver Architectures}
\label{system_model_section}
\subsection{RSMA system model}
 Consider a Base Station (BS), equipped with $N_t$ transmit antennas, that serves $K$ downlink single-antenna communication users, indexed by the set $\mathcal{K} = \{1,\dots,K\}$. We consider 1-layer RSMA as described in \cite{eurasip} which uses a single common stream for any number of users. In RSMA, the message of user-$k$, $W_k$, is split into a common part $W_{c,k}$ and a private part $W_{p,k}$, $\forall k\in\mathcal{K}$. Then, the common parts of all $K$ users $\{W_{c,1},\dots,W_{c,K}\}$ are jointly encoded and modulated into a single common stream $s_c$, while the private parts $\{W_{p,1},\dots,W_{p,K}\}$ are encoded and modulated separately into $K$ private streams $\{s_1,\dots,s_K\}$. Although non-linearly precoded RSMA has been investigated recently \cite{lina_dpc, lamare_thp}, linearly precoded RSMA remains the predominant case of study in the literature \cite{perfect_1,perfect_3,perfect_4,perfect_5,joudeh_2,imperfect_1,imperfect_2,imperfect_3,imperfect_4,imperfect_5,imperfect_6}. In the latter, the data streams are linearly precoded using the precoder $\mathbf{P} = [\mathbf{p}_c,\mathbf{p}_1,\dots,\mathbf{p}_K] \in \mathbb{C}^{N_t \times (K+1)}$, where $\mathbf{p}_c$ is the common stream precoder and $\mathbf{p}_k$ is the private stream precoder for user-$k$. The transmitted signal $\mathbf{x} \in \mathbb{C}^{N_t \times 1}$, subject to the transmit power constraint $\mathbb{E}\{||\mathbf{x}||^2\}\leq P_t$, is then given by
\begin{equation}
    \mathbf{x} = \mathbf{P}\mathbf{s} = \mathbf{p}_cs_c + \sum_{k=1}^{K}\mathbf{p}_ks_k,
    \label{rsma_transmit_signal}
\end{equation}
where $\mathbf{s} = [s_c,s_1,\dots,s_K]^T \in \mathbb{C}^{(K+1)\times 1}$. It is assumed that $\mathbb{E}\{\mathbf{s}\mathbf{s}^H\}=\mathbf{I}_{(K+1)}$. Hence, $\Tr(\mathbf{P}\mathbf{P}^H)\leq P_t$. At user-$k$, the received signal at the output of the antenna is given by
\begin{equation}
    \begin{split}
        y_k &= \mathbf{h}_k^H\mathbf{P}\mathbf{s} + n_k \\
        &= \mathbf{h}_k^H\mathbf{p}_cs_c+ \mathbf{h}_k^H\mathbf{p}_k s_k+\overbrace{\sum_{j\neq k,j\in\mathcal{K}}\mathbf{h}_k^H\mathbf{p}_j s_j}^{\text{MUI}} + n_k,
    \end{split}
    \label{rsma_eq}
\end{equation}
where $\mathbf{h}_k \in \mathbb{C}^{N_t \times 1}$ is the channel between the transmitter and user-$k$, and $n_k\;\mathtt{\sim}\; \mathcal{CN}(0,\sigma_{n,k}^2)$ is the Additive White Gaussian Noise (AWGN) at user-$k$.

The CSI model is given by \cite{lina_dpc}
\begin{equation}
    \mathbf{H}=\hat{\mathbf{H}}+\Tilde{\mathbf{H}},
\end{equation}
where $\mathbf{H}=[\mathbf{h}_1,\dots,\mathbf{h}_K]$ is the real CSI with the entries of $\mathbf{h}_k$ being i.i.d complex Gaussian entries drawn from the distribution $\mathcal{CN}(0,\sigma_k^2), \forall k\in \mathcal{K}$, and $\sigma_k^2$ being the channel amplitude power. Also, $\hat{\mathbf{H}}=[\hat{\mathbf{h}}_1,\dots,\hat{\mathbf{h}}_K]$ is the estimated CSIT/CSIR with $\hat{\mathbf{h}}_k$ following a Gaussian distribution $\mathcal{CN}(0,\sigma_k^2-\sigma_{e,k}^2), \forall k\in \mathcal{K}$. Finally, $\Tilde{\mathbf{H}}=[\Tilde{\mathbf{h}}_1,\dots,\Tilde{\mathbf{h}}_K]$ represents the CSI estimation error in the CSI estimation/acquisition process shown in Fig. \ref{fig:rsma_system_model}, with $\Tilde{\mathbf{h}}_k$ following a Gaussian distribution $\mathcal{CN}(0,\sigma_{e,k}^2), \forall k\in \mathcal{K}$. The parameter $\sigma_{e,k}^2$ is defined as the CSIT error power for user-$k$. The perfect CSIT scenario can then be represented by choosing $\sigma_{e,k}^2 = 0$. 

\subsection{RSMA receiver architectures}
In this subsection, receiver architectures to enable RSMA communications are presented.

\subsubsection{Maximum a-posteriori probability receiver}
The optimum receiver is the maximum a-posteriori probability (MAP) receiver, which performs joint detection and decoding of the common and private streams, in the case of 1-Layer RSMA, for user-$k$ while minimizing the decoded bit error rate \cite{map_1}. The MAP receiver outputs an estimation of the common and private message pair $\hat{W}_c, \hat{W}_{p,k}$ that maximizes the a-posteriori probability $ p(W_c, W_{p,k}|y_k)$ over all possible message pairs of the common and private sets $\mathcal{W}_c, \mathcal{W}_{p,k}$. The MAP estimation can then be expressed as follows by applying Bayes' rule
\begin{equation}
\begin{split}
\hat{W}_c, \hat{W}_k &\triangleq \argmax _{W_c \in \mathcal{W}_c, W_{p,k} \in \mathcal{W}_{p,k}}  p(W_c, W_{p,k}|y_k)\\
&=\argmax _{W_c \in \mathcal{W}_c, W_{p,k} \in \mathcal{W}_{p,k}} \frac{p(W_c, W_{p,k}) p(y_k|W_c, W_{p,k})}{p(y_k)}.
\end{split}
\label{map_rule}
\end{equation}

The first term in the numerator in (\ref{map_rule}) is the a-priori probability of the transmitted bit sequence pair and the second term is the likelihood function. The denominator term is a normalization factor which can ultimately be omitted \cite{map_rule}. If the transmitted bit sequence pairs are assumed to be equiprobable, then the first term can also be omitted  and (\ref{map_rule}) turns into the maximum likelihood (ML) rule. If a further assumption is made so that the channel is considered a linear Gaussian channel, then (\ref{map_rule}) turns into minimum distance estimation, which can be expressed in closed form as 
{
\begin{equation}
\begin{split}
  \hat{W}_c, \hat{W}_k &= \argmin _{W_c \in \mathcal{W}_c, W_k \in \mathcal{W}_k} ||y_k -\mathbf{h}_k^H\mathbf{p}_c\phi_c(W_c)-\mathbf{h}_k^H\mathbf{p}_k\phi_k(W_k) ||^2,
    \label{ml_joint_gaussian}
    \end{split}
\end{equation}}
\normalsize
where $\phi_c(.), \phi_k(.)$ denote the encoding and modulation function of the common and private stream of user-$k$, respectively. Then, $\hat{W}_c$ is splitted to recover $\hat{W}_{c,k}$ and it is combined with $\hat{W}_{p,k}$ to reconstruct the message $\hat{W}_{k}$.
\subsubsection{Successive interference cancellation receiver}

\begin{figure}[t!]
		\centering
		\includegraphics[width=0.6\textwidth]{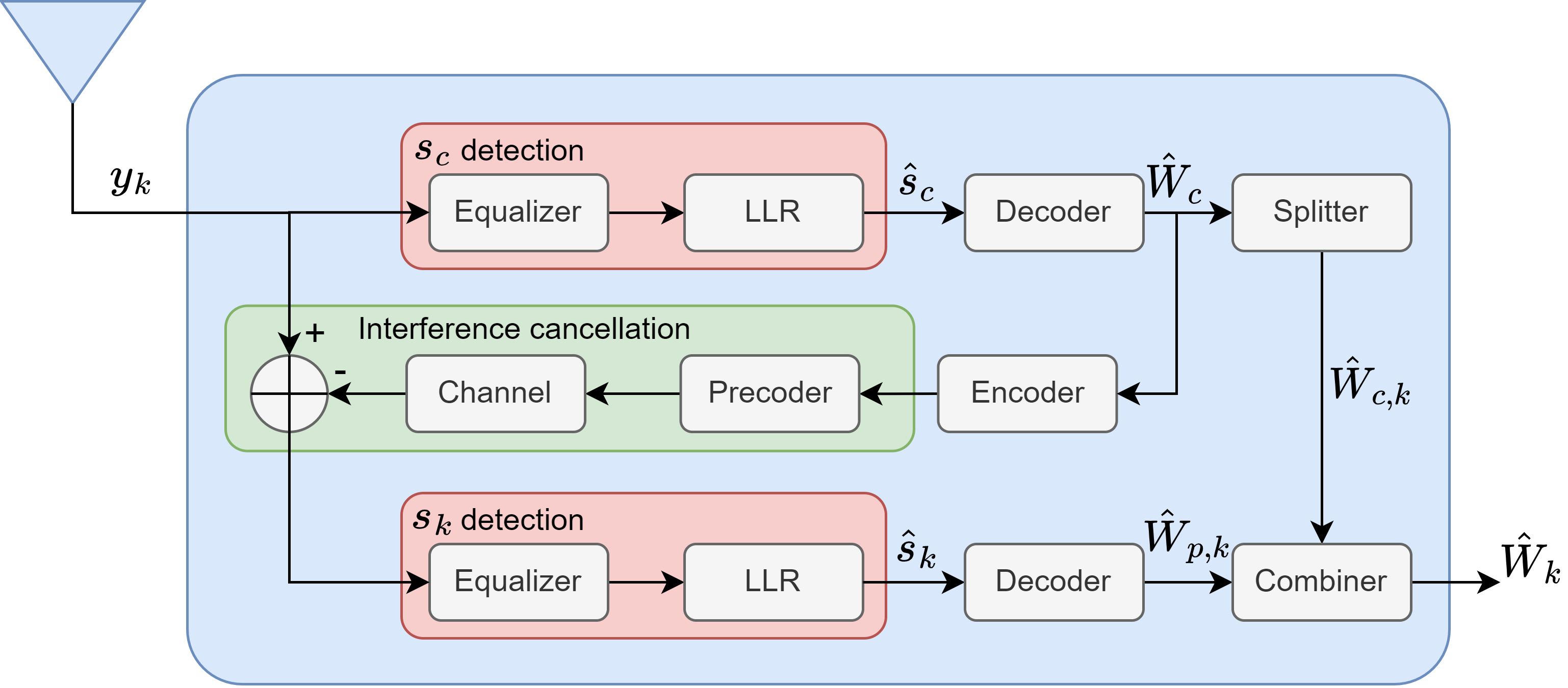}
		\caption{1-Layer SIC RSMA receiver architecture \cite{rs_overview}.}
		\label{fig:rsma_receiver_model}
\end{figure}

This receiver architecture, depicted in Fig. \ref{fig:rsma_receiver_model}, is the main one used in past works. By applying SIC, it performs successive decoding of the common and private streams with affordable complexity compared to the optimum MAP receiver \cite{eurasip}. Its operation is described next.

First, the common stream $s_c$ is decoded into $\hat{W}_c$ by treating the interference from the $K$ private streams as noise. The Signal-to-Interference-and-Noise Ratio (SINR) of decoding $s_c$ is given by
\begin{equation}
    \gamma_{c,k} = \frac{|\mathbf{h}_k^H\mathbf{p}_c|^2}{\sum_{k\in\mathcal{K}}|\mathbf{h}_k^H\mathbf{p}_k|^2+\sigma_{n,k}^2}.
    \label{commonsinr}
\end{equation}
Then, assuming perfect CSIR, $\hat{W}_c$ is re-encoded, precoded, multiplied by the channel vector and subtracted from $y_k$ using SIC so that the private stream $s_k$ can be decoded into $\hat{W}_{p,k}$ by treating the remaining $K-1$ private streams as noise. The SINR of decoding $s_k$ is then given by
\begin{equation}
    \gamma_{p,k} = \frac{|\mathbf{h}_k^H\mathbf{p}_k|^2}{\sum_{j\neq k,j\in\mathcal{K}}|\mathbf{h}_k^H\mathbf{p}_j|^2+\sigma_{n,k}^2}.
    \label{privatesinr}
\end{equation} 

Finally, user-$k$ extracts $\hat{W}_{c,k}$ from $\hat{W}_c$ and combines it with $\hat{W}_{p,k}$ to reconstruct the message $\hat{W}_k$. Therefore, the achievable rate of the common stream for user-$k$ is  $R_{c,k} = \log_2(1+\gamma_{c,k})$ and the achievable rate of its corresponding private stream is $R_{k} = \log_2(1+\gamma_{k})$. To guarantee that all $K$ users are able to decode the common stream successfully, it must be transmitted at a rate that does not exceed $R_c     = \min\{R_{c,1},\dots,R_{c,K}\}$.

\normalsize

\normalsize
\section{Model-Based Deep Learning Rate-Splitting Multiple Access Receiver Design}
As highlighted in the previous section, the less studied area of past RS research efforts has been the design and optimization of a practical receiver with imperfect CSIR. Due to the underlying potential of DL in wireless communication applications, as multi-core processing becomes available in state-of-the-art hardware \cite{dl_harware}, it is first proposed that a DL-based RS receiver is designed. 

\subsection{Model-based deep learning}
An intuitive approach to a practical receiver design is then to take simultaneous advantage of the tractability and simplicity of model-based algorithms, and the model-agnosticism of DNNs through \textit{model-based deep learning} techniques \cite{model-based-dl,model-based-dl-2}.

The purpose of both model-based and DL-based techniques is to perform inference. That is, they generate a prediction of a labeled variable $\hat{\mathbf{s}}\in\mathcal{S}$ based on an input variable $\mathbf{x}\in\mathcal{X}$, where $\mathcal{S}$ is named the \textit{label space}, and $\mathcal{X}$, the \textit{input space} \cite{model-based-dl-2}. However, model-based methods rely mainly on \textit{domain knowledge} $p_{\mathbf{x}|\mathbf{s}}$, while DL-based methods depend on the fidelity of available training data $\{\mathbf{x}_t,\mathbf{s}_t\}$. Thus, these two types of techniques represent the two extremes of the domain-knowledge/data spectrum shown in Fig. \ref{fig:data_know_spectrum}.

On the other hand, model-based deep learning (MBDL) systems are hybrid model-based/data-driven systems tailored for specific applications that can be trained with small data sets and operate without a full prior knowledge of the underlying model. Thus, they are located in the middle region of the domain-knowledge/data spectrum. Depending on their design method, MBDL systems can be divided in the following two classes \cite{model-based-dl}:
\begin{enumerate}
    \item \textbf{Model-aided networks:} This strategy focuses on using the structure of model-based algorithm as a template to design a customized DNN architecture for the application of interest. In order to achieve this, the steps from the model-based algorithms that require full model knowledge are first identified. Then, dedicated DNNs, each of them less complex than the DNN that would be used in a conventional DL system, are designed and tuned to replace the identified model-based steps.
\begin{figure*}[t]
		\centering
		\includegraphics[width=\textwidth]{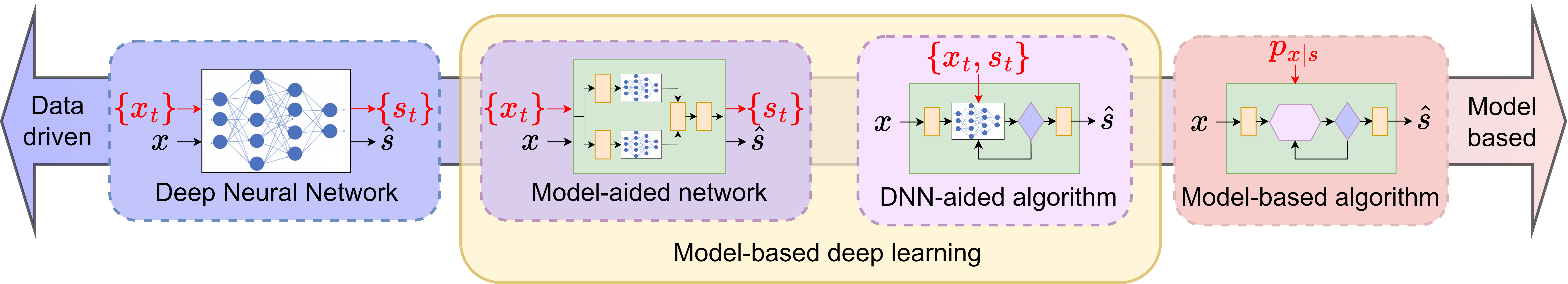}
		\caption{Domain-knowledge/data spectrum \cite{model-based-dl}.}
		\label{fig:data_know_spectrum}
\end{figure*}
    The resulting network can then be trained in an end-to-end manner to jointly optimize all of the dedicated DNNs at the expense of requiring a sufficiently large training set. Another method is to train the dedicated DNNs in a sequential manner, which would decrease the total training data set size but may increase the training time as some DNNs may need the output of previous ones as inputs.
    
    \item \textbf{DNN-aided algorithms:} This strategy aims to empower the operation of conventional model-based algorithms by replacing specific computations that require prior model knowledge with compact dedicated DNNs. As these are only introduced to perform certain calculations and not to generate a new DNN structure, they can be separately trained from the system in an offline manner. Also, due to their small size and limited complexity, the required training data sets are smaller than those of model-aided networks.
\end{enumerate}

 \begin{figure}[t]
		\centering
		\includegraphics[width=\textwidth]{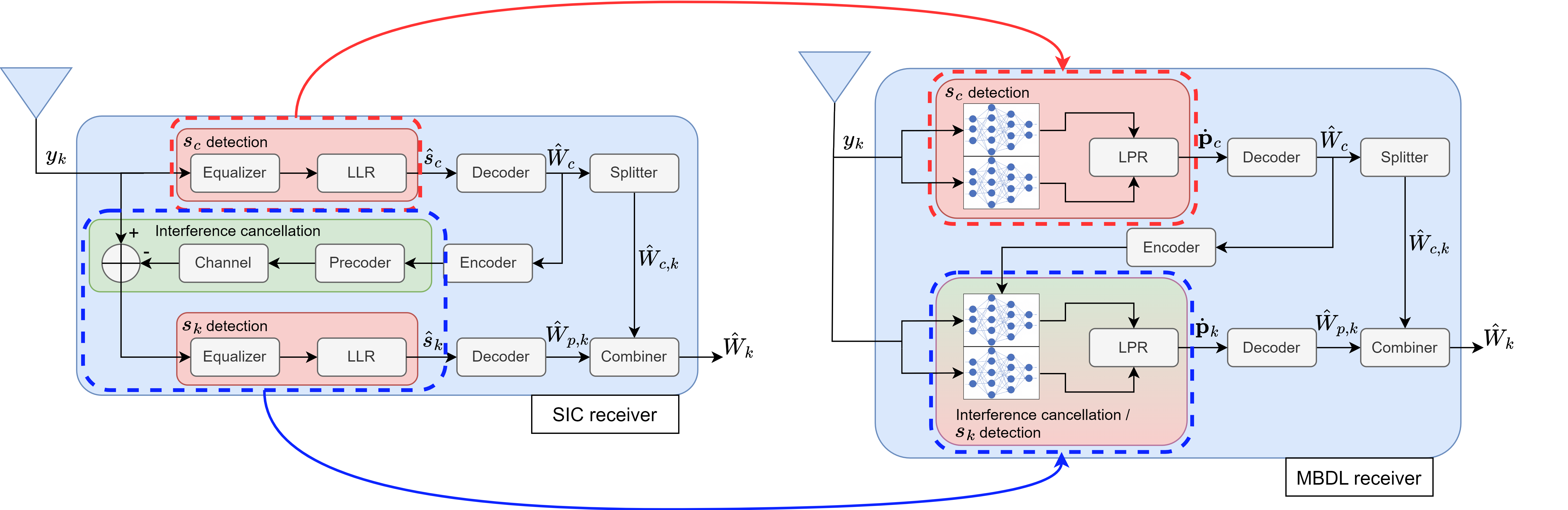}
		\caption{1-Layer MBDL RSMA receiver and relationship with 1-Layer SIC RSMA receiver.}
		\label{fig:model-based-dl-rs-receiver}
\end{figure}
\subsection{Model-based deep learning rate-splitting receiver architecture}
In this subsection, we describe the proposed MBDL-based RSMA receiver adapted from the SIC receiver structure. Since the filtering, detection and interference cancellation tasks of the SIC receiver require CSIR (prior model knowledge), these modules can be replaced with dedicated DNNs in a model-aided-network approach, as shown in Fig. \ref{fig:model-based-dl-rs-receiver}, where, for each stream, a bank of two DNNs is employed at user-$k$ to jointly generate the soft symbol bit estimate of the common stream $\mathbf{\dot{p}}_{c} \in \mathbb{R}_+^{M_c \times 1}$, where $|\mathcal{S}_{c}|=2^{M_c}$ is the common stream modulation order and $M_c$ is the number of bits in each common stream symbol, and the soft symbol bit estimate of the private stream $\mathbf{\dot{p}}_{k} \in \mathbb{R}_+^{M_k \times 1}$, where $|\mathcal{S}_{k}|=2^{M_k}$ is the modulation order of the $k$-th private stream, and $M_k$ is the number of bits in each symbol of the $k$-th private stream. The architecture of each of these DNNs is summarized in Table \ref{dnn_architecture}. The details of the proposed algorithm are as follows:

\begin{table}[t!]
\centering
\caption{DNNs architecture of 1-Layer MBDL RSMA receiver.}
\label{dnn_architecture}
\begin{tabular}{|c|c|c|c|cccc|}
\hline
       \multirow{2}{*}{Purpose} & \multirow{2}{*}{Layer} & \multirow{2}{*}{Type} & \multirow{2}{*}{\makecell{\text{Activation}\\\text{function}}} & \multicolumn{4}{c|}{\# Neurons (for target constellation)}                                                    \\ \cline{5-8} 
                                        &                   &                   &                   & \multicolumn{1}{c|}{QPSK} & \multicolumn{1}{c|}{16QAM} & \multicolumn{1}{c|}{64QAM} & 256QAM \\ \hline
\multicolumn{1}{|c|}{\multirow{5}{*}{\makecell{$s_c$ detection}}} &          Input         &        ---           &     ---              & \multicolumn{4}{c|}{2} \\ \cline{2-8} 
\multicolumn{1}{|c|}{}                  &        Hidden           &         FC          &        Sigmoid           &   \multicolumn{1}{c|}{10} & \multicolumn{1}{c|}{15} & \multicolumn{1}{c|}{20} & 25 \\ \cline{2-8} 
\multicolumn{1}{|c|}{}                  &       Hidden            &         FC         &        ReLU           & \multicolumn{1}{c|}{10} & \multicolumn{1}{c|}{15} & \multicolumn{1}{c|}{20} & 25 \\ \cline{2-8} 
\multicolumn{1}{|c|}{}                  &       Hidden            &         FC          &       ReLU            & \multicolumn{1}{c|}{---} & \multicolumn{1}{c|}{---} & \multicolumn{1}{c|}{---} & 25 \\ \cline{2-8} 
\multicolumn{1}{|c|}{}                  &       Output            &         FC         &        Softmax           & \multicolumn{1}{c|}{2} & \multicolumn{1}{c|}{4} & \multicolumn{1}{c|}{8} & 16 \\ \hline  \hline
\multicolumn{1}{|c|}{\multirow{5}{*}{ \makecell{Interference \\ cancellation / \\ $s_k$ detection} }} &       Input         &        ---           &     ---              & \multicolumn{4}{c|}{2+$M_c$} \\ \cline{2-8} 
\multicolumn{1}{|c|}{}                  &       Hidden            &         FC          &       Sigmoid            & \multicolumn{1}{c|}{20} & \multicolumn{1}{c|}{25} & \multicolumn{1}{c|}{30} & 35 \\ \cline{2-8} 
\multicolumn{1}{|c|}{}                  &       Hidden            &         FC          &       ReLU            & \multicolumn{1}{c|}{20} & \multicolumn{1}{c|}{25} & \multicolumn{1}{c|}{30} & 35 \\ \cline{2-8} 
\multicolumn{1}{|c|}{}                  &       Hidden            &         FC          &       ReLU            & \multicolumn{1}{c|}{---} & \multicolumn{1}{c|}{---} & \multicolumn{1}{c|}{---}  & 35 \\ \cline{2-8} 
\multicolumn{1}{|c|}{}                  &       Output            &         FC          &       Softmax             & \multicolumn{1}{c|}{2} & \multicolumn{1}{c|}{4} & \multicolumn{1}{c|}{8} & 16 \\ \hline
\end{tabular}
\end{table}

\begin{itemize}
    \item The received signal $y_k $ is divided into its real and imaginary parts to form the vector $[\Re\{y_k\}, \Im\{y_k\}]^T \in \mathbb{R}^{2\times 1}$, which is used as input for the DNNs used in the detection of $s_c$. After decoding and re-encoding the common stream message, the input vector of the DNNs for interference cancellation and the detection of $s_k$ is given by $[\Re\{y_k\}, \Im\{y_k\},\hat{\mathbf{b}}_c^T]^T\in \mathbb{R}^{(2+M_c)\times 1}$, where $\hat{\mathbf{b}}_c^T\in\mathbb{R}^{M_c\times 1}$ is the estimated symbol bit vector of $s_c$.
    \item \textit{Fully connected} (FC) layers are employed. This means that the inputs of all neurons in the FC layer are all the neuron outputs in the previous layer. 
    \item The non-linear activation functions sigmoid, $S(x)=\frac{1}{1+e^{-x}}$, and ReLU, $R(x)=\max(0,x)$, are employed after the hidden layer outputs in order to allow the DNN to approximate non-linear detection boundaries.
    \item The purpose of employing two DNNs in each bank is to detect each stream by estimating the row and column of the modulation constellation, in which the symbol of interest falls in. In this way, it is possible to simplify the complex classification problem that would be imposed if a single DNN was used. To illustrate this, consider the classification of 256QAM symbols. In the case that a single DNN is used for classification, 256 different classes would have to be learned. Thus, the training data must at least be composed of 256 different symbols. With the proposed architecture, however, only 16 different classes are learned by each DNN, as the 256QAM constellation has 16 rows/columns. Consequently, the training set must be composed of at least 16 different symbols and a significant reduction in the training set size and DNN complexity can be achieved.
    \item The cross-entropy is commonly used as the loss function in classification problems. Therefore, in order to guarantee a suitable output probability vector $\hat{\mathbf{p}} = \in \mathbb{R}_+^{d \times 1}$, where $d$ is the size of the output layer, so that the cross-entropy loss calculation can be valid (i.e. $\hat{\mathbf{p}} \geq \mathbf{0}$ and $\sum_{i=1}^{d}\hat{p}_i = 1$), a \textit{softmax} activation function is used after after the output layer. This activation function is defined by 
    \begin{equation*}
        \text{Softmax}(\mathbf{x})=\Bigg<\frac{\exp(x_1)}{\sum_{i=1}^d \exp(x_i)},\dots,\frac{\exp(x_d)}{\sum_{i=1}^d \exp(x_i)}\Bigg>.
    \end{equation*}
    \item For a symbol with $M$ bits, the $M/2$ left-most bits denote in which column the symbol is located in the constellation; and the $M/2$ right-most bits, in which row. Thus, the m-$th$ soft symbol bit estimate in $\mathbf{\dot{p}}_{c}$ or $\mathbf{\dot{p}}_{k}$ can be obtained by
    \begin{equation}
        \dot{p}_m=\sum_{d \in  \mathcal{D}_m^1} \hat{p}_d,
    \end{equation}
    where $\mathcal{D}_m^1$ is the subset of row/column indices in the modulation constellation for which the m-$th$ bit is equal to `1'. Then, the bit log-probability ratio (LPR) is calculated as  \cite{demodnet}
    \begin{equation}
        \text{LPR}(\dot{p}_m)=\log\Big(\frac{1-\dot{p}_m}{\dot{p}_m}\Big)\;,\;m=1,2,\dots,M.
    \end{equation}
    The LPRs can then be used as substitutes for the bit log-likelihood ratios (LLRs) that are used in conventional channel decoders.
\end{itemize}
\subsection{Training symbol set pattern design for the MBDL receiver}
A key performance indicator in any practical wireless receiver is the system training overhead $\overline{T}$ given by
\begin{equation}
    \overline{T}=\frac{T}{L}100\%,
    \label{overhead_eq}
\end{equation}
where $T$ is the number of training symbols in the training set $\{\mathbf{x}_t,\mathbf{s}_t\}$ and $L$ is the total number of transmitted symbols (including training and data symbols) assuming block fading, in which the coherence time of the fading process is equal or larger than the required time to transmit the $L$ total symbols. Using conventional channel estimation methods, $T$ must be at least equal to the number of unknown parameters to estimate  \cite{channel_estimation}, which would be $T \geq 1$ for the single-antenna receivers in the 1-Layer RSMA system model. Naturally, there exists a trade-off between the CSIR quality and $\overline{T}$. This trade-off is particularly critical for DL-based receivers, as classical DL techniques require a large $\{\mathbf{x}_t,\mathbf{s}_t\}$ to appropriately learn the optimum channel mapping function $f_{\bm{\theta}}(\mathbf{x})$. Hence the importance of an optimized MBDL architecture also designed to minimize $T$.

The issue of $\overline{T}$ in DL-based receivers is not commonly addressed in the literature, as past works usually consider that the $\{\mathbf{x}_t,\mathbf{s}_t\}$ is available prior to transmission (\textit{offline training}) \cite{dl_offline_1,dl_offline_2}, a large $\{\mathbf{x}_t,\mathbf{s}_t\}$ can be transmitted (\textit{online training})  \cite{noma_dl_2,dl_online_2}, or that $\{\mathbf{x}_t,\mathbf{s}_t\}$ can be generated at the receiver based on $\hat{\mathbf{h}}_k$  \cite{dl_estimation_1}.
However, as $\mathbf{h}_k$ would still change for future blocks, the DNNs parameters must be updated to track these channel variations, rendering these approaches impractical. For this reason, the following training set design methods are proposed to deploy the MBDL receiver in block fading channels, with the basic building unit of the training set being named \textit{training sub-block}.

\subsubsection{Extensive Training}
Under this design method, the training block is composed by superposing all possible symbol combinations in the $(K+1)$ data streams. In other words, the training block contains $|\mathcal{S}_c||\mathcal{S}_1|\dots|\mathcal{S}_K|$ superposed training symbols. Thus, this approach allows the DNNs in each receiver to learn the full pattern of the MUI. However, as the size of the training block increases with $K$ and the modulation order of each data stream, it is obvious that this training pattern is suitable only when $K$ is small, low order modulation is employed, or when very slow block fading is experienced.

\subsubsection{Minimal Training}
This training pattern is designed to take advantage of the fact that RS precoders are optimized to minimize MUI to noise level. Using this method, the $K$ private streams are first arranged in descending order according to their respective modulation orders. Then, the basic training block is constructed by superposing all possible symbol combinations between $\mathcal{S}_c$ and $\mathcal{S}_1$, and random combinations of the symbols in the remaining $(K-1)$ private streams. In this way, at least one copy of every symbol combination between the common stream and each private stream is transmitted to every user. The training block of the minimal training effectively contains $|\mathcal{S}_c||\mathcal{S}_1|$ superposed training symbols, which is independent of $K$ but still scales with the modulation orders of $\mathcal{S}_c$ and $\mathcal{S}_1$. 

 \begin{figure*}[t!]
\begin{minipage}{.5\linewidth}
\centering
\subfloat[]{\label{main:interpolation_a}\includegraphics[scale=.55]{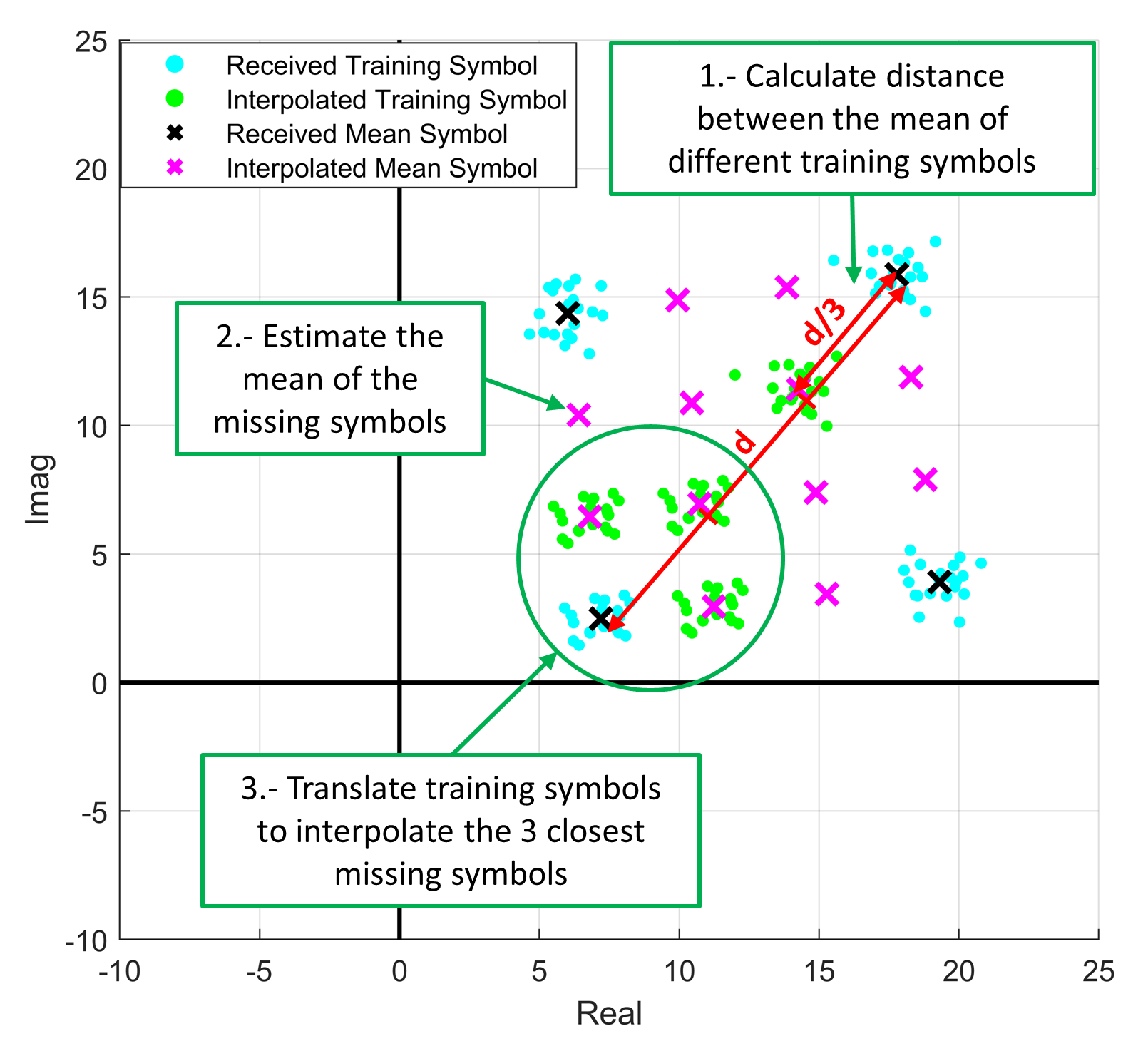}}
\end{minipage}%
\begin{minipage}{.5\linewidth}
\centering
\subfloat[]{\label{main:interpolation_b}\includegraphics[scale=.55]{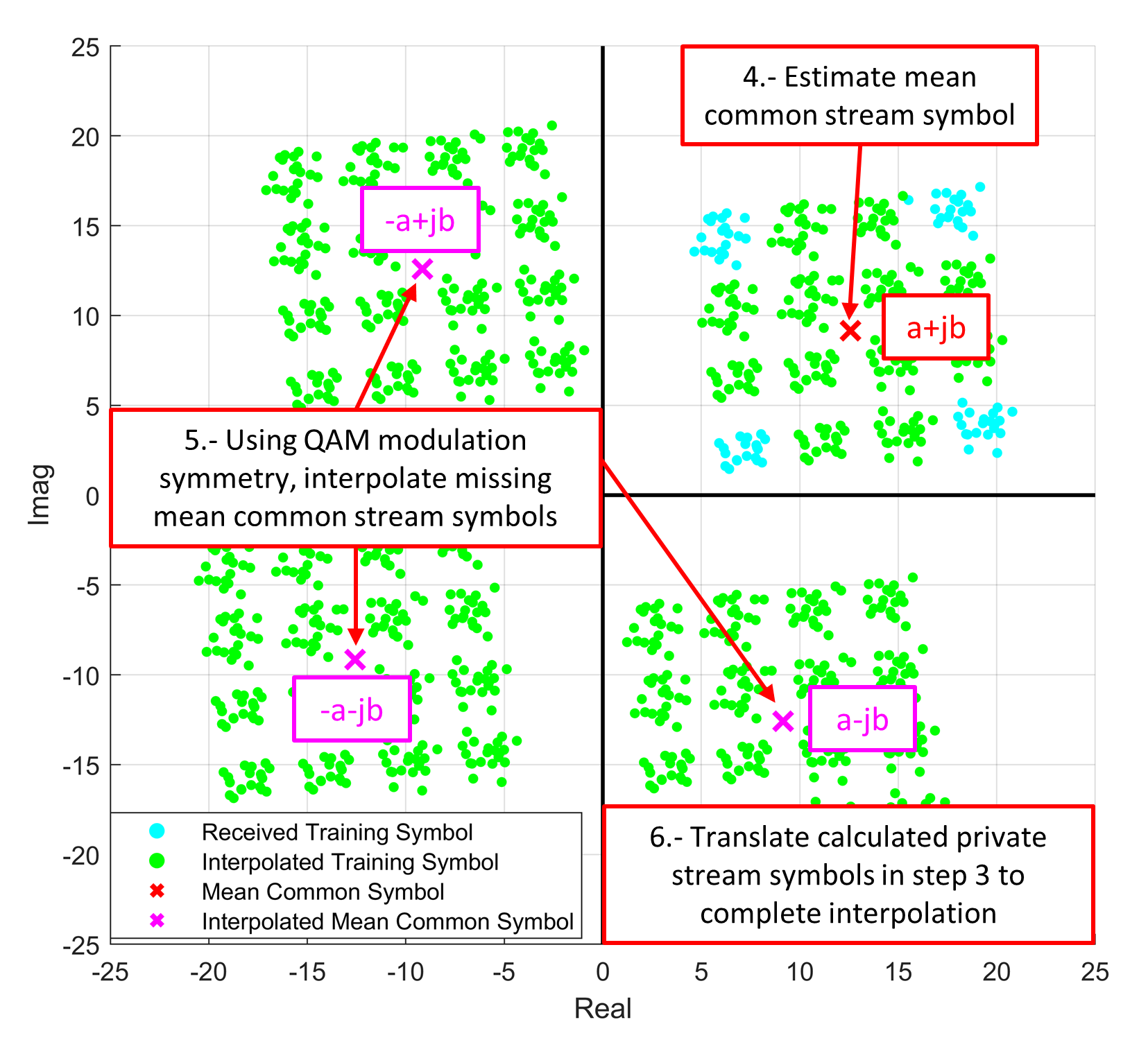}}
\end{minipage}\par\medskip
\caption{Interpolation training example: (a) Private stream symbols ($\mathcal{S}_1=$ 16QAM) (b) Common stream symbols ($\mathcal{S}_c=$ QPSK).}
\label{fig:interpolation}
\end{figure*}
\subsubsection{Interpolating Training}
This scheme takes advantage of the symmetry of QAM constellations to further make the size of each training block independent of $K$ and the modulation order of any data stream. This is achieved by superposing only the symbols located in the corners of the QAM constellations of the $K$ private streams in a randomized manner in each training block. Thus, each training block contains only 4 superposed training symbols. At the receiver of user-$k$, the missing training symbols are interpolated in a similar manner as in the example shown in Fig. \ref{fig:interpolation}. However, it is worth indicating that this approach assumes that the channel is linear so that the QAM constellation symmetry is maintained in the received signal $y_k$.
\subsection{Computational complexity analysis}
The complexity of fully-connected DNNs can be measured in terms of the number of trainable parameters \cite{springer_dl} and the number of real multiplications per symbol (RMpS) \cite{dl_metric_1}. The complexity of the DNNs for each modulation type and each type is then presented in Table \ref{dnn_complexity_table}.
\begin{table}[h]
\caption{DNN complexity comparison}
\resizebox{\columnwidth}{!}{\begin{tabular}{|c|c|cccc|}
\hline
\multirow{2}{*}{\makecell[t]{Complexity \\ metric}} & \multirow{2}{*}{DNN Type} & \multicolumn{4}{c|}{Target constellation}                                                    \\ \cline{3-6} 
                  &                   & \multicolumn{1}{c|}{QPSK} & \multicolumn{1}{c|}{16QAM} & \multicolumn{1}{c|}{64QAM} & 256QAM \\ \hline
\multirow{2}{*}{\makecell{\# Trainable \\ parameters}} &   $s_0$ detection                & \multicolumn{1}{c|}{$162$} & \multicolumn{1}{c|}{$349$} & \multicolumn{1}{c|}{$648$} & $1791$ \\ \cline{2-6} 
                  &           \makecell{Interference\\ cancellation /\\$s_k$ detection}        & \multicolumn{1}{c|}{$522+20M_c$} & \multicolumn{1}{c|}{$829+25M_c$ } & \multicolumn{1}{c|}{$1268+30M_c$} & $3181+35M_c$ \\ \hline
\multirow{2}{*}{\makecell{\# Real \\multiplications \\ per symbol \\ (RMpS)}} &         $s_0$ detection            & \multicolumn{1}{c|}{$140$} & \multicolumn{1}{c|}{$315$} & \multicolumn{1}{c|}{$600$} & $1700$ \\ \cline{2-6} 
                  &          \makecell{Interference\\ cancellation /\\$s_k$ detection}         & \multicolumn{1}{c|}{$480+20M_c$} & \multicolumn{1}{c|}{$775+25M_c$ } & \multicolumn{1}{c|}{$1200+30M_c$} & $3080+35M_c$  \\ \hline
\end{tabular}}
\label{dnn_complexity_table}
\end{table}

It is worth noticing that the largest RMpS is in the order of $3\times10^3$, which makes the DNNs in this work feasible for a practical DNN hardware implementation \cite{dl_metric_1}.

\section{Simulation Results}

In this section, we present the performance of the MBDL receiver in terms of uncoded SER, throughput through LLS tests, and training overhead, averaged over 500 random channel realizations, and compare it with the model-based MAP and SIC receivers with perfect and imperfect CSIR. MATLAB is used to run all simulations and its Deep Learning toolbox is used to construct and train the DNNs of the MBDL receiver.
\subsection{Simulation setup}
\begin{figure*}[t]
		\centering
		\includegraphics[width=\textwidth]{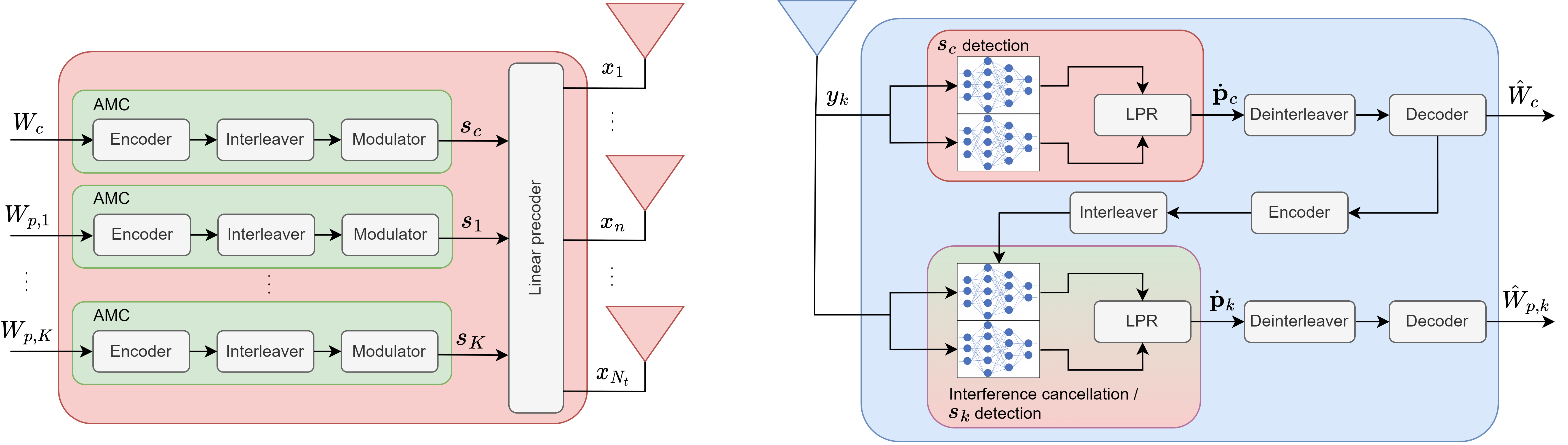}
		\caption{1-Layer RSMA PHY-layer architecture with MBDL receivers (adapted from \cite{onur_lls}).}
		\label{fig:lls_arch}
\end{figure*}

Unless otherwise stated, the parameters used in the simulations are listed in Table \ref{simulation_table}. 
To obtain uncoded SER results, uncoded data streams are employed, each carrying $10^5$ symbols. In this way, the detection capabilities of the DNNs in the MBDL receiver can be directly analyzed without the error correction aid of practical channel coding. Also, only imperfect CSIT and perfect CSIR is considered. In turn, to calculate the throughput we employ the 1-Layer RSMA PHY-layer architecture shown in Fig. \ref{fig:lls_arch}, which features finite alphabet modulation, finite-length polar coding \cite{polar_code}, and an Adaptive Modulation and Coding (AMC) algorithm. The function of the AMC algorithm is to select an appropriate modulation-coding rate pair based on the transmit rate calculation obtained from the precoder optimization. The throughput is then defined as
\begin{table}[t!]
    \centering
    \caption{Simulation parameters.}
    \label{simulation_table}
    \resizebox{0.5\columnwidth}{!}{%
    \begin{tabular}{|c|c|}
    \hline
    \textbf{Parameter} & \textbf{Value} \\
    \hline
     \# Transmit antennas & $N_t=\{4,8,16\}$ \\
     \# Communication users & $K=8$ \\
     Fading channel model & Rayleigh \\
     User channel variance & $\sigma_k^2=1, \forall k \in \mathcal{K}$ \\
     CSIT/CSIR error estimation variance & $\sigma_{e,k}^2=P_t^{-\alpha}, \forall k \in \mathcal{K}$\\
     CSI quality scaling factor & $\alpha=0.6$ \\
     Noise variance at the receivers & $\sigma_{n,k}^2=1, \forall k \in \mathcal{K}$ \\
     Precoder optimization method & \makecell{SAA-WMMSE-AO \\ algorithm [2]} \\
     DNN optimizer & Adam \\
     DNN learning rate & $0.01$ \\
     \# Training epochs $(s_c \text{ detection})$ & $12.5\log_2(|\mathcal{S}_c|)$ \\
     \makecell{\# Training epochs (interference \\ cancellation / $s_k$ detection)}& $12.5\log_2(|\mathcal{S}_k|)$ \\
     DNN mini-batch size & $\max(T,25|\mathcal{S}_1|)$ \\
     \hline
    \end{tabular}
    }
    \end{table}

\begin{equation}
    \text{Throughput [bps/Hz]} = \frac{\sum_l \sum_k  D_{s,k}^{(l)}}{\sum_l S^{(l)}},
\end{equation}
where $S^{(l)}$ denotes the modulated block length in the $l$-th Monte Carlo channel realization, and $D_{s,k}^{(l)}$ denotes the number of received bits by user-$k$ in the common stream (considering only its intended part of the common message) and private stream when there are no decoding errors in a given block. In all simulations, $S^{(l)} = 256$ is used. Also, rate backoff is introduced to compensate the loss due to finite-length coding, bit interleaving scheme, and imperfect CSIT and imperfect CSIR. A MAP decoding bound is also included in the throughput results to highlight the loss incurred by a practical Successive Cancellation List (SCL) polar decoder. This bound is calculated following the indications in \cite{scl}. Finally, the calculation of the training overhead is done assuming the worst case scenario that only one modulated block length of 256 symbols can be transmitted without the channel changing due to the effects of the fading process. 

\subsection{Symbol error rate evaluation}
\label{ser_results}

\begin{figure}[t!]
    \centering
    \begin{minipage}{0.47\linewidth}
    \centering
    \subfloat[Common stream: QPSK. Private stream: QPSK.]{\label{main:a_ser}\includegraphics[scale=.59]{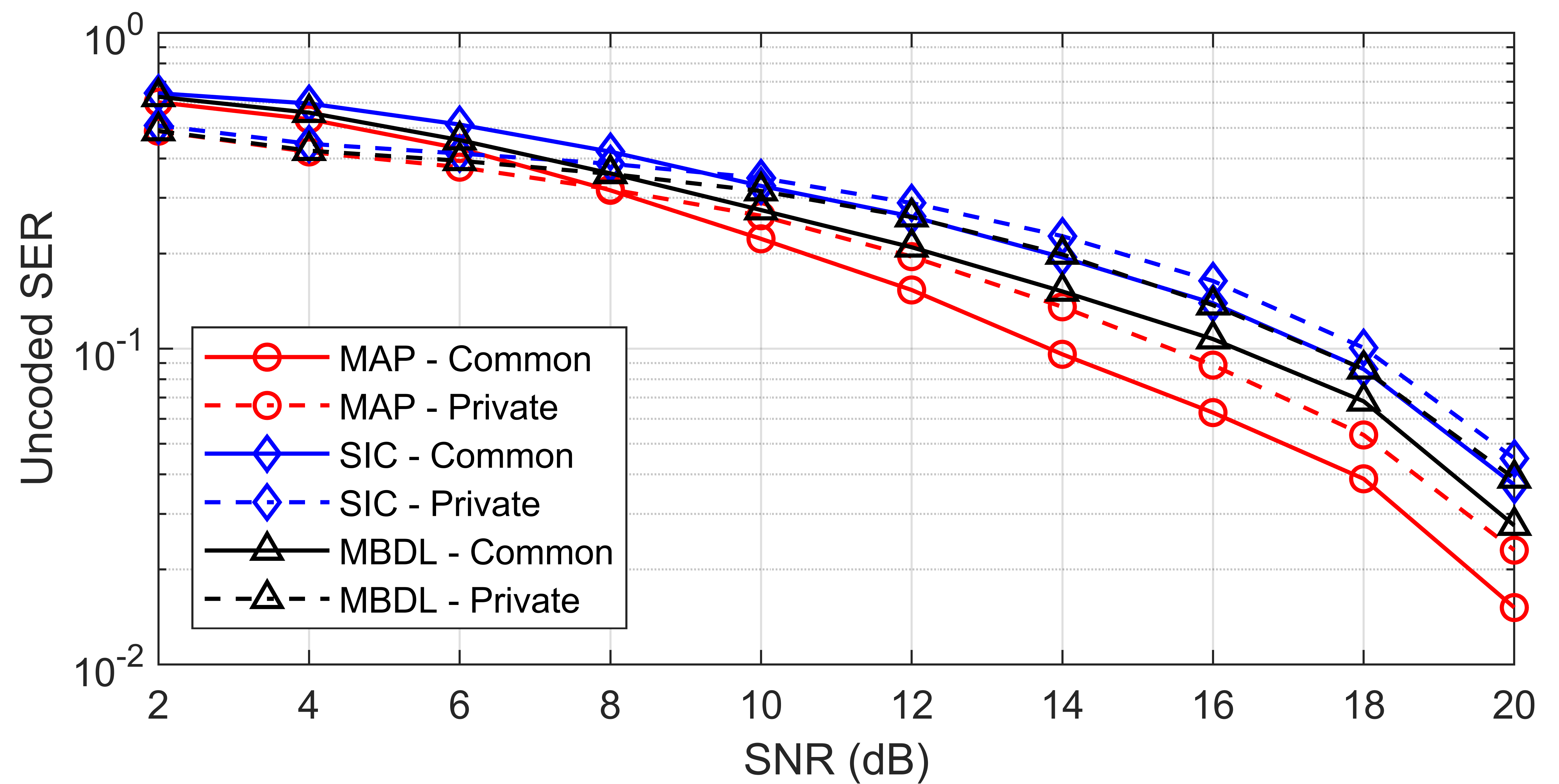}}
    \end{minipage}
    \begin{minipage}{0.47\linewidth}
    \centering
    \subfloat[Common stream: QPSK. Private stream: 16QAM.]{\label{main:b_ser}\includegraphics[scale=.59]{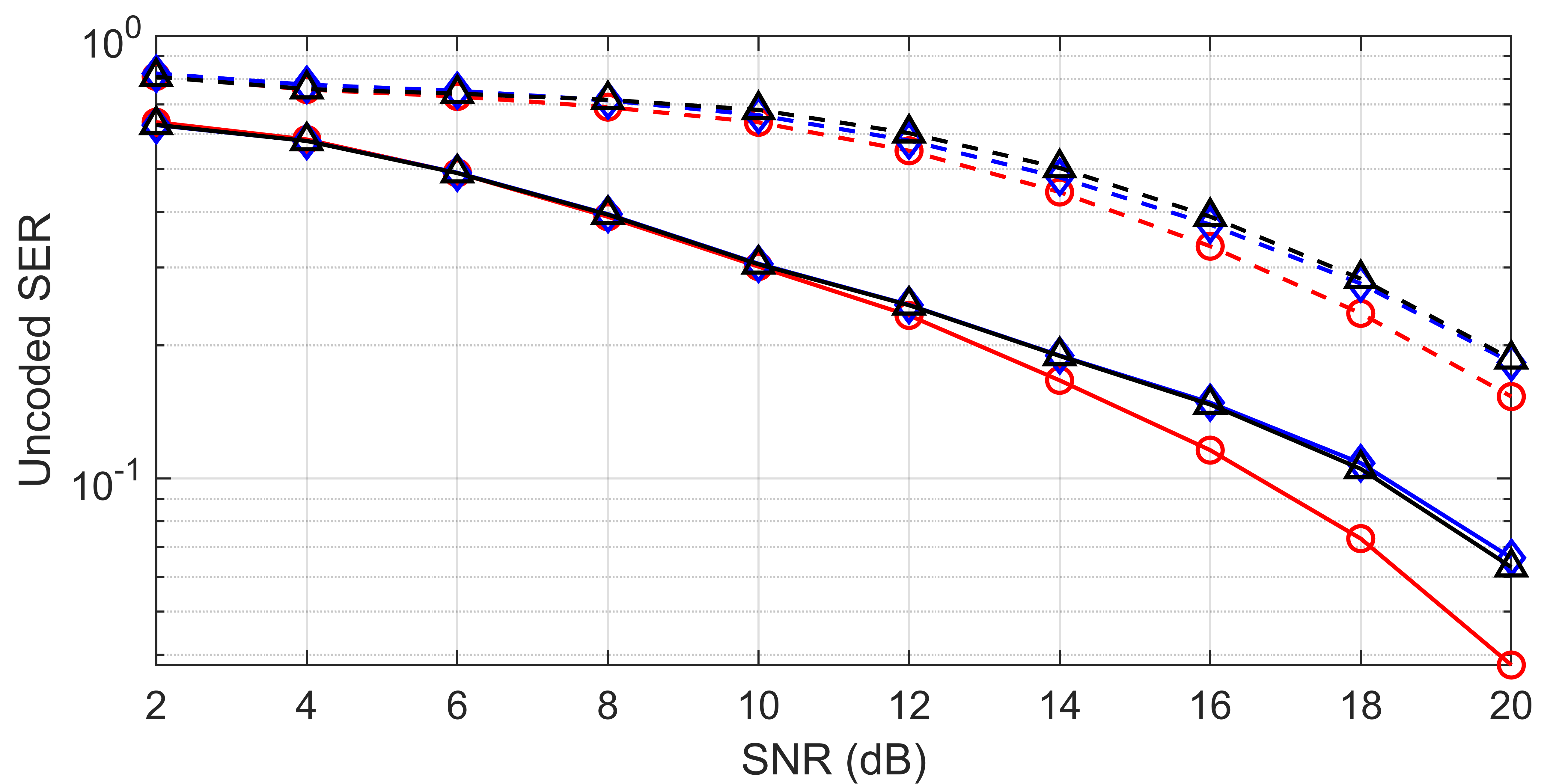}}
    \end{minipage}\par\medskip
    \begin{minipage}{0.47\linewidth}
    \centering
    \subfloat[Common stream: QPSK. Private stream: 64QAM/256QAM.]{\label{main:c_ser}\includegraphics[scale=.59]{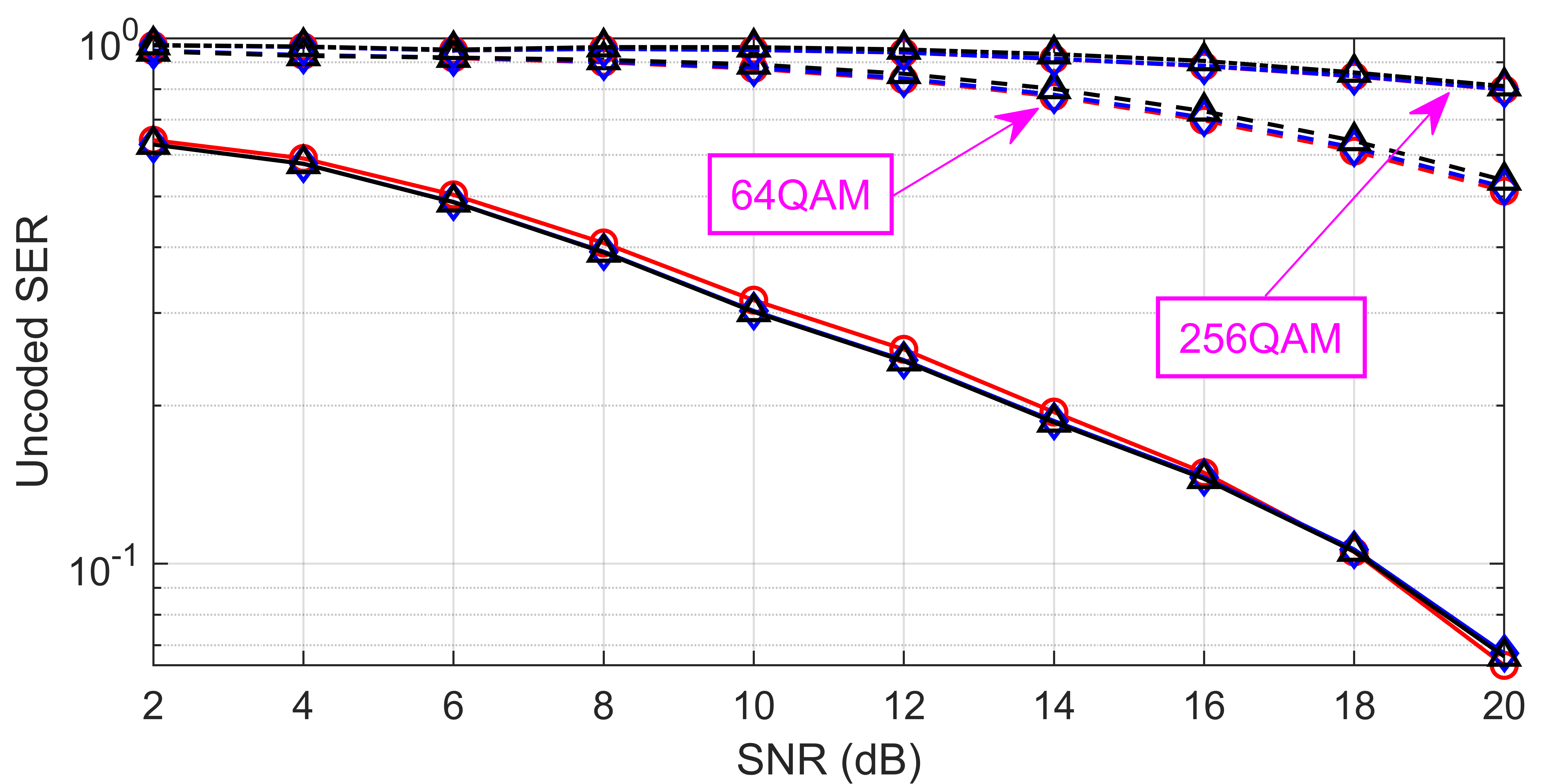}}
    \end{minipage}\par\medskip
    \caption{Uncoded SER vs. SNR (minimal training with 20 training blocks).}
    \label{fig:ser_comparison}
\end{figure}

\begin{figure*}[t!]
   \centering
	\includegraphics[width=0.45\textwidth]{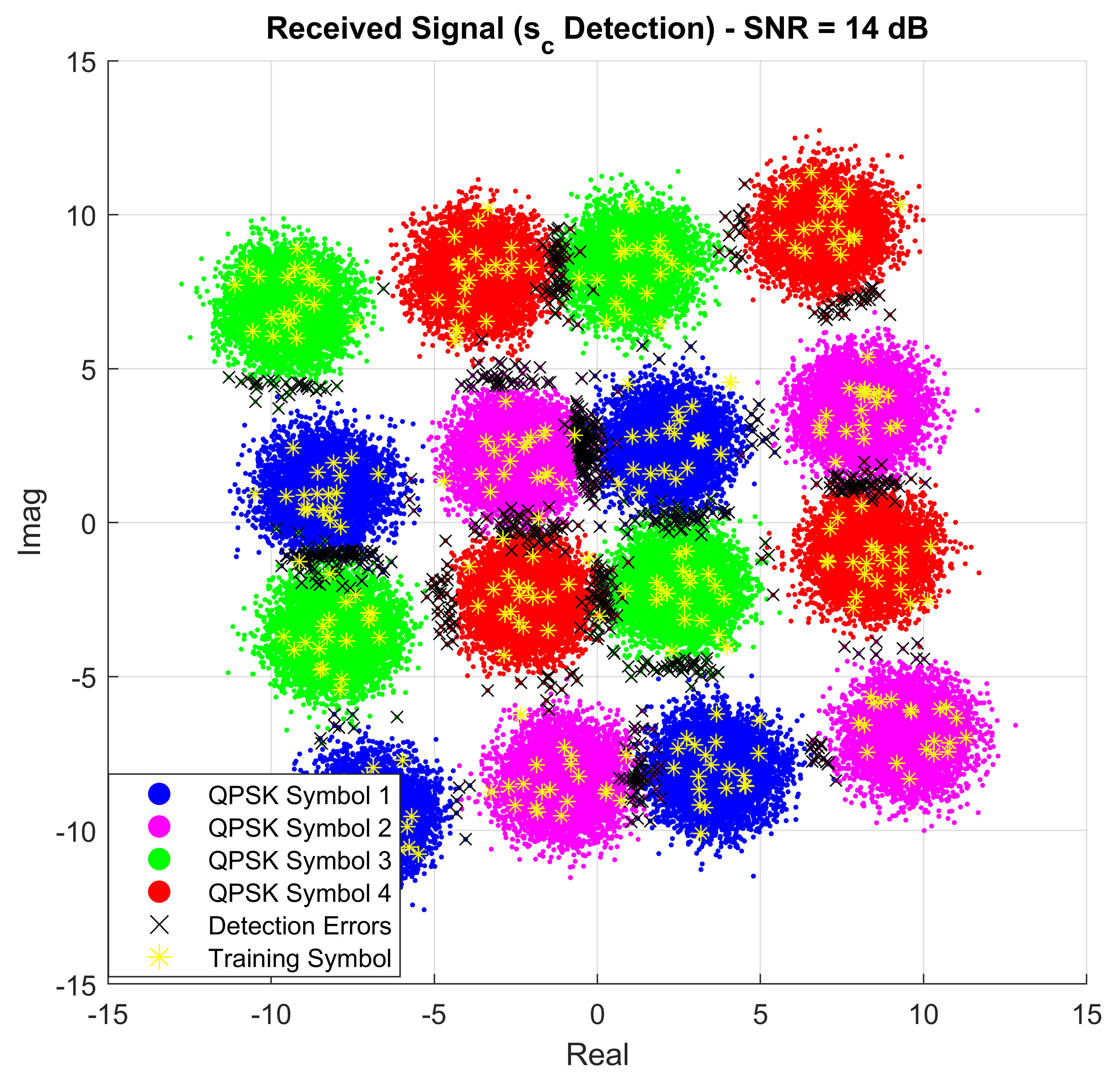}
	\caption{Non-linear detection of QPSK common stream symbols (minimal training with 20 training blocks).}
	\label{fig:non_linear_common}
\end{figure*}

The average SER results, assuming $N_t=4$ and $K=2$,
achieved by user-1 are then plotted in Fig. \ref{fig:ser_comparison} for different orders of the private stream modulation. It is then observed from Fig. \ref{main:a_ser}, where QPSK is used for both the common and private streams, that the MBDL receiver outperforms the SIC receiver for SNR levels higher than 6 dB, and also approaches the SER achieved by the optimum MAP receiver (the latter boils down to the ML receiver as all symbols have the same probability of being transmitted). This gain over the SIC receiver is due to the ability of the MBDL receiver to generate non-linear decision boundaries for symbol detection, whereas the SIC receiver is only able to use the linear decision boundaries dictated by the real and imaginary axes after signal equalization. This effect is shown in Fig. \ref{fig:non_linear_common}, where well-differentiated, but not linearly separable, symbol clusters can be observed. By employing its non-linear detection boundaries, the MBDL receiver confines the symbol detection errors only to regions where the symbol clusters corresponding to different common stream symbols overlap or where there is not enough training data available. Thus, this minimizes error propagation and facilitates the detection of the private stream symbols. Regarding the MAP receiver, it is noticed that it slightly outperforms the MBDL receiver, as the latter generates non-linear decision boundaries based solely on the available 320 training symbols in the 20 training blocks (TBs), which does not fully capture the effects of the channel, MUI and noise that the $10^5$ test symbols experience. Thus, to close this gap more training symbols can be transmitted at the cost of increasing the training overhead.

As the modulation order of the private stream increases, so does the SER of the common and private streams, as the chosen modulation schemes may not be suitable for the given SNR and the precoders have also not been optimized to minimize the SER for finite size modulation schemes. Nevertheless, it is still seen in Fig. \ref{main:b_ser} and Fig. \ref{main:c_ser} that MBDL still performs similarly to the MAP and SIC receivers even in the presence of large adjacent-symbol interference when using 16QAM, 64QAM and 256QAM to modulate the private stream.

\begin{figure*}[tp]
\begin{minipage}{0.47\linewidth}
\centering
\subfloat[Minimal training: $N_t=4, K=8$.]{\label{main:throughput_4_8_minimal}\includegraphics[width=\textwidth]{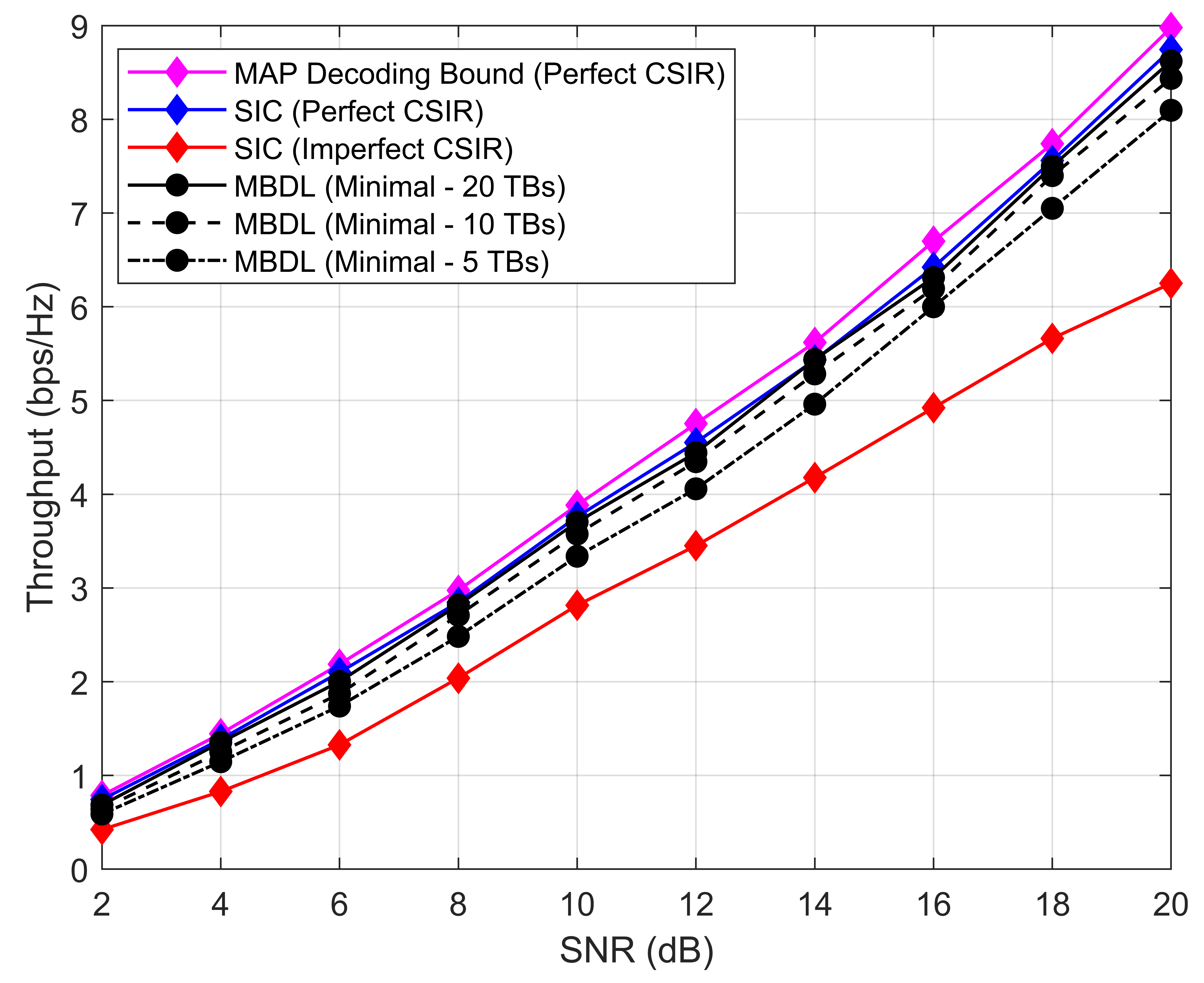}} 
\end{minipage}
\begin{minipage}{0.47\linewidth}
\centering
\subfloat[Interpolating training: $N_t=4, K=8$.]{\label{main:throughput_4_8_interpolating}\includegraphics[width=\textwidth]{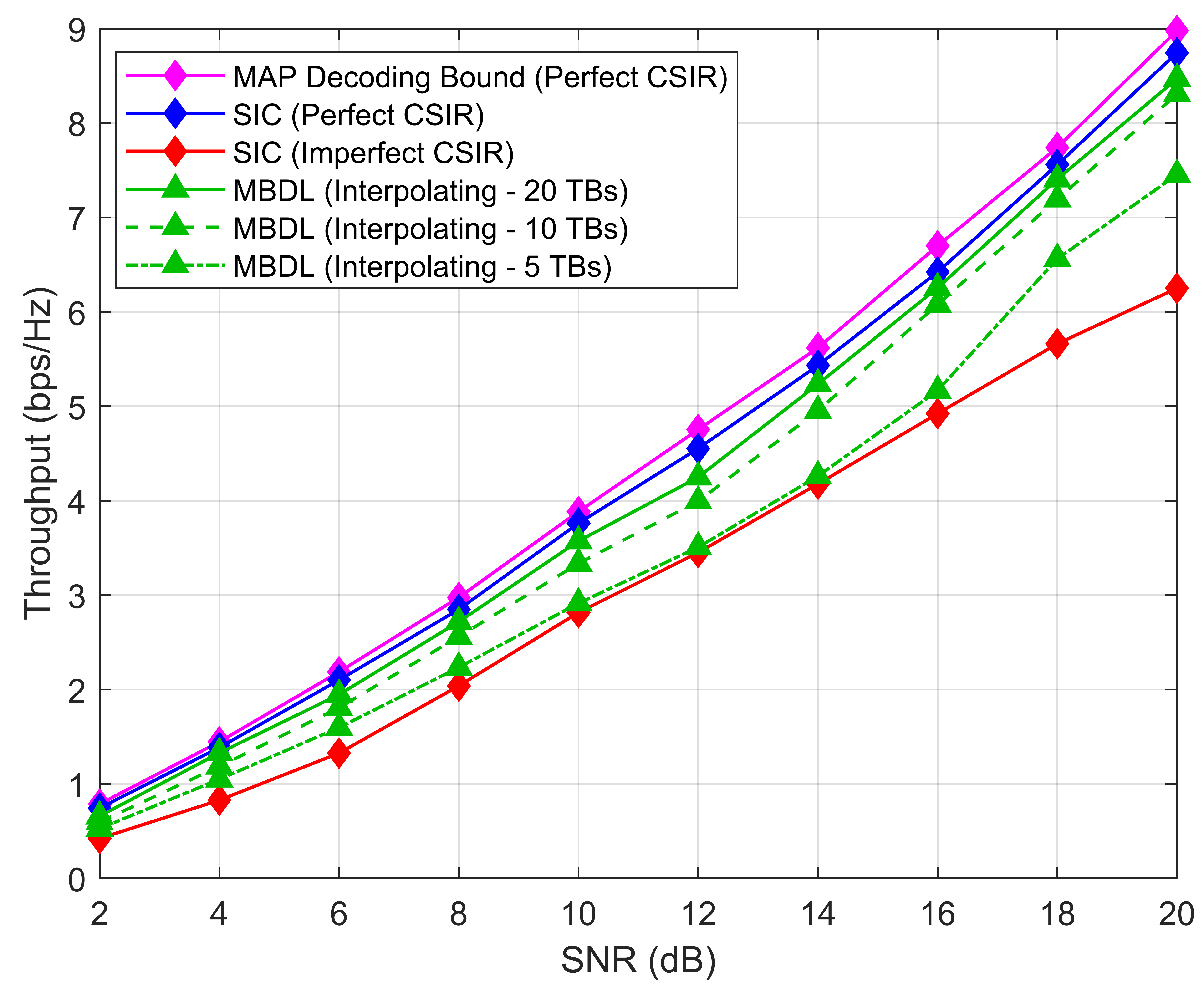}}
\end{minipage}\par\medskip
\begin{minipage}{0.47\linewidth}
\centering
\subfloat[Minimal training: $N_t=8, K=8$.]{\label{main:throughput_8_8_minimal}\includegraphics[width=\textwidth]{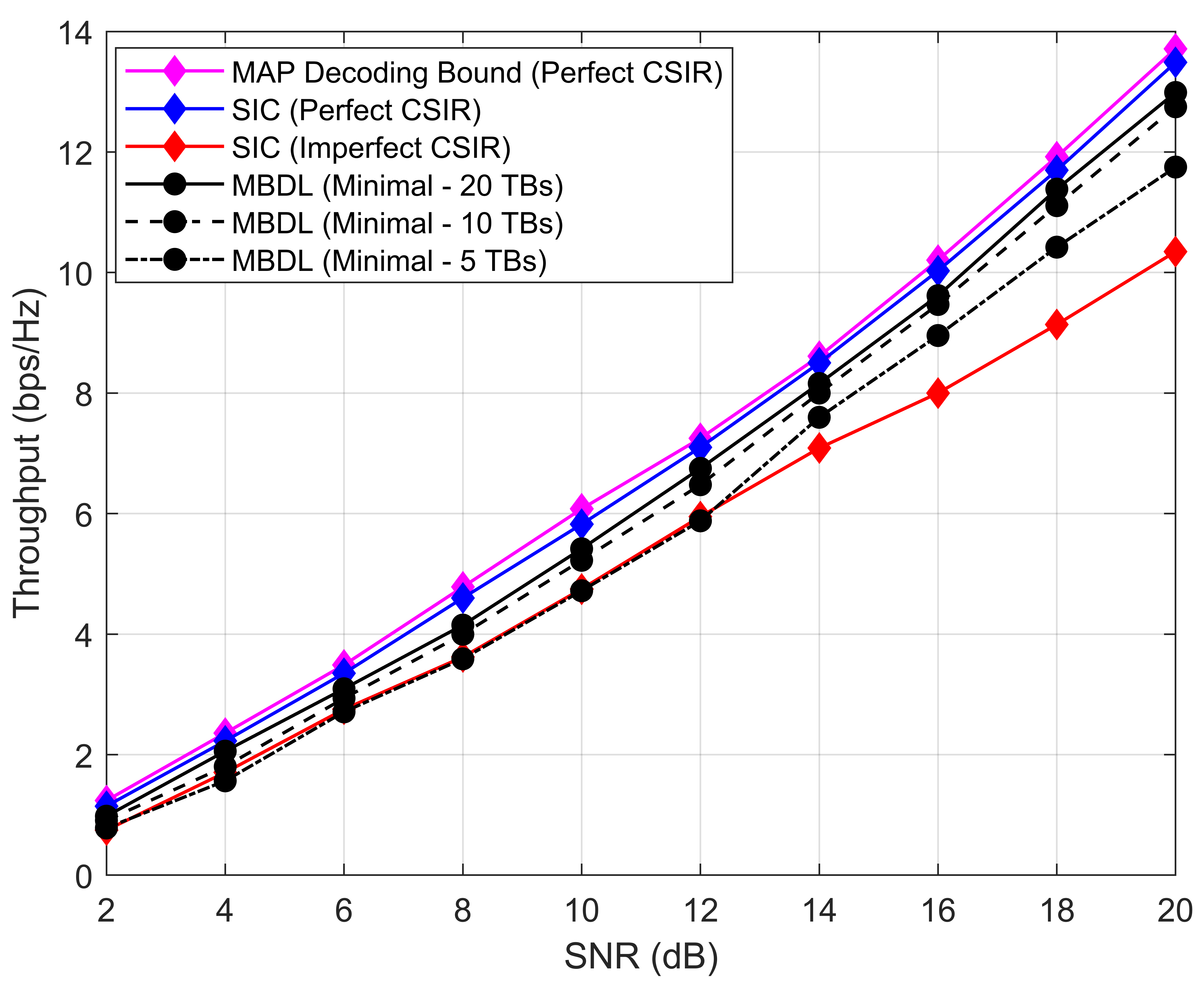}} 
\end{minipage}
\begin{minipage}{0.47\linewidth}
\centering
\subfloat[Interpolating training: $N_t=8, K=8$.]{\label{main:throughput_8_8_interpolating}\includegraphics[width=\textwidth]{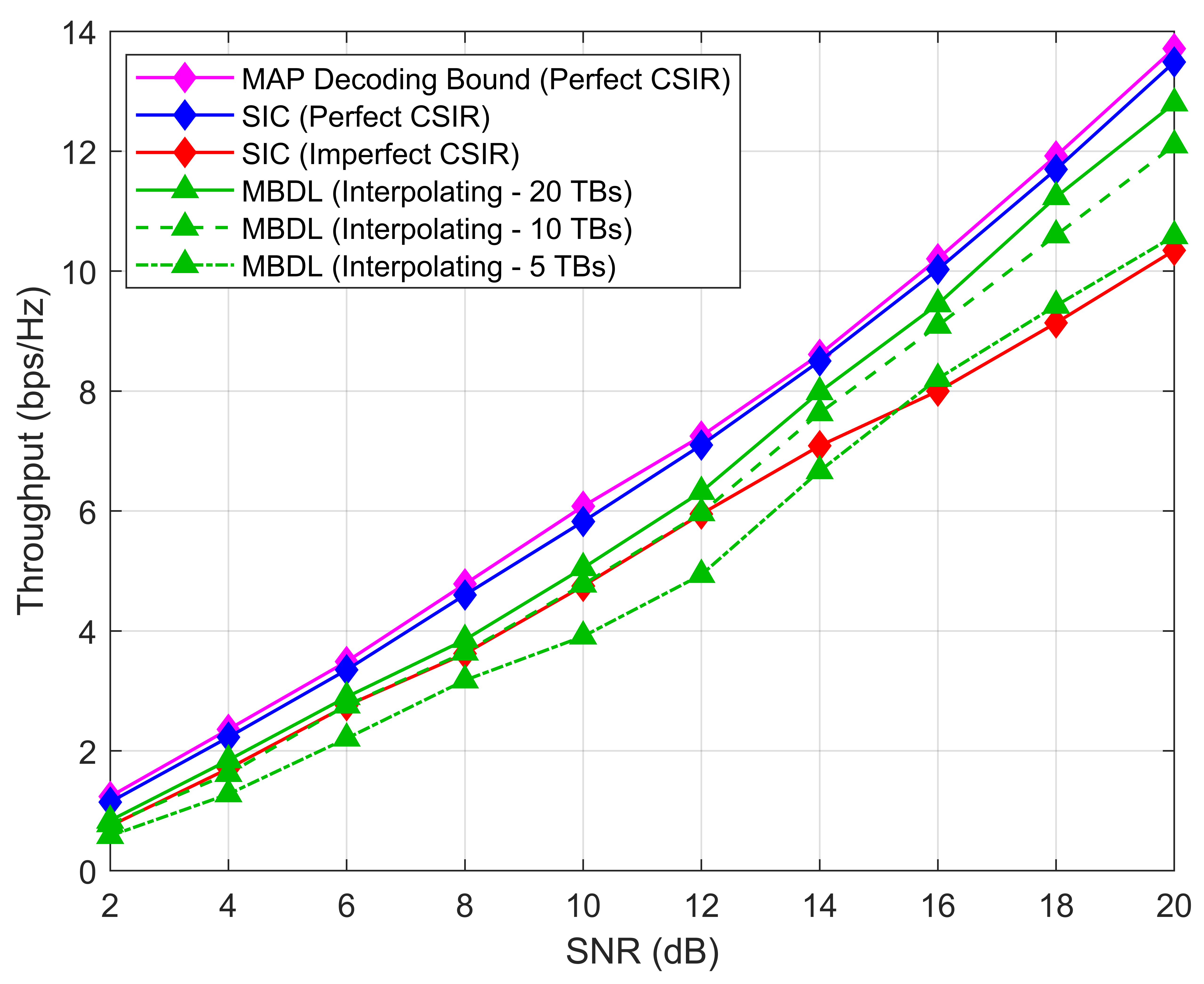}}
\end{minipage}\par\medskip
\begin{minipage}{0.47\linewidth}
\centering
\subfloat[Minimal training: $N_t=16, K=8$.]{\label{main:throughput_16_8_minimal}\includegraphics[width=\textwidth]{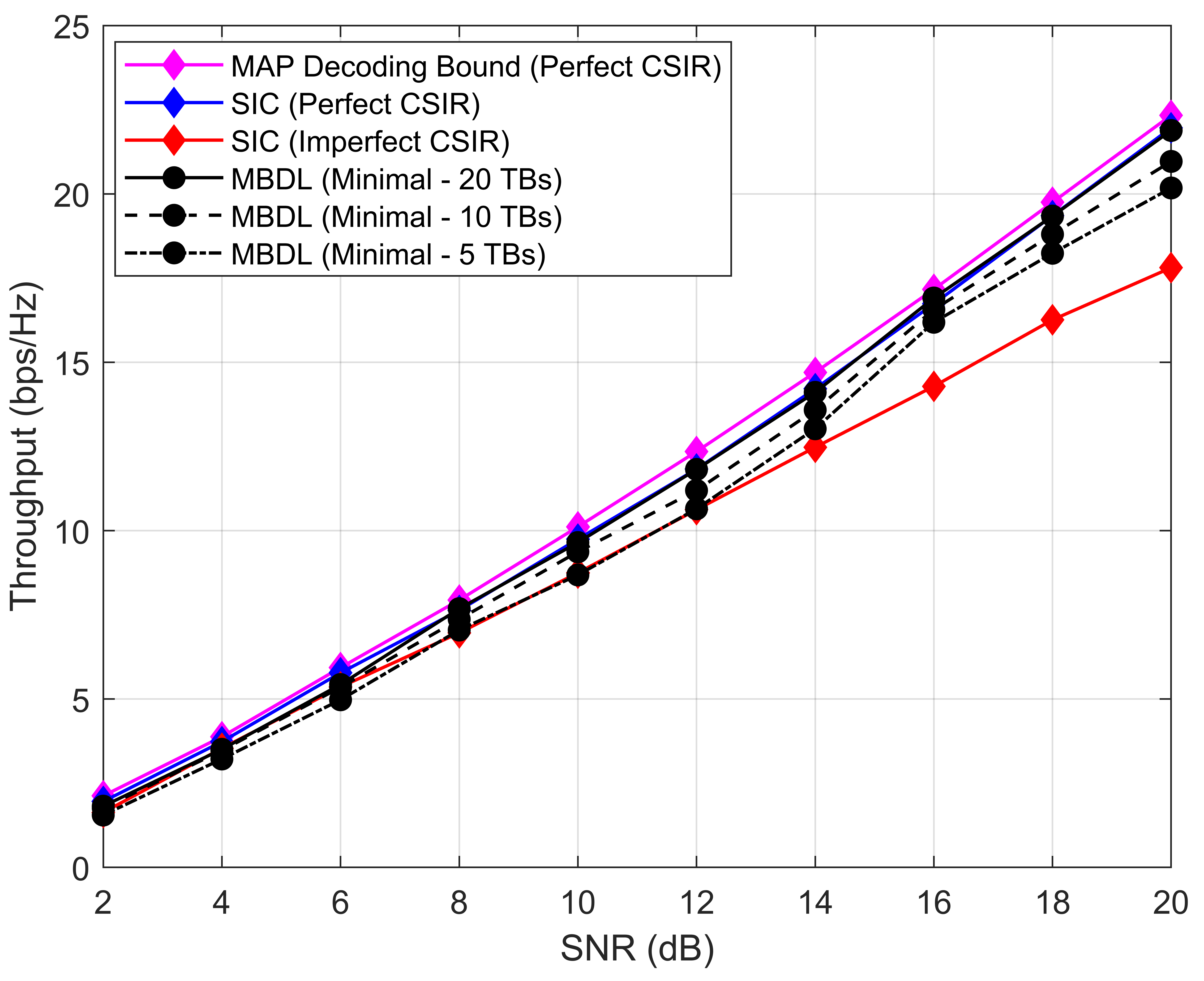}} 
\end{minipage}
\begin{minipage}{0.47\linewidth}
\centering
\subfloat[Interpolating training: $N_t=16, K=8$.]{\label{main:throughput_16_8_interpolating}\includegraphics[width=\textwidth]{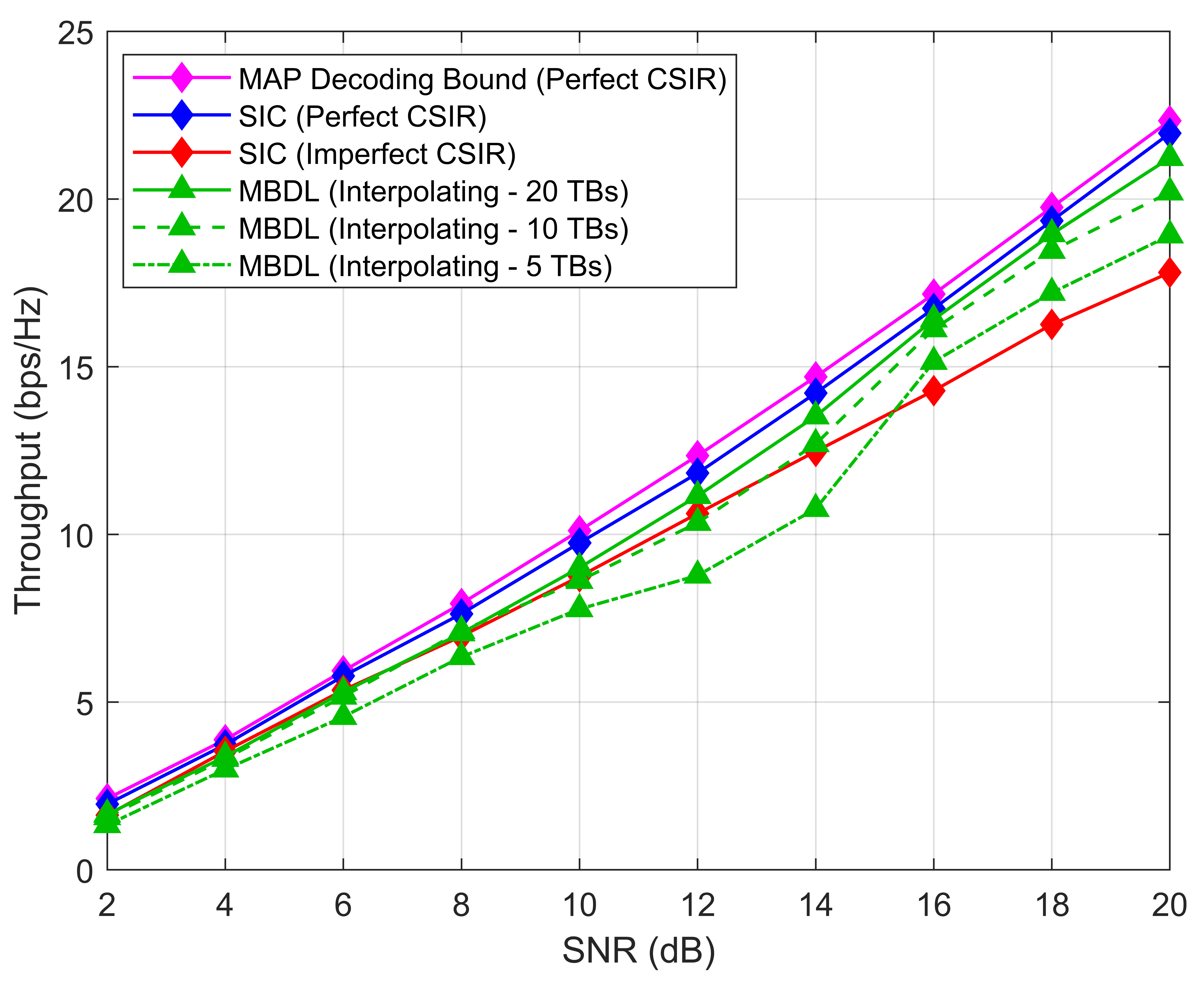}} 
\end{minipage}\par\medskip
\caption{Throughput vs. SNR.}
\label{fig:throughput}
\end{figure*}

\subsection{Throughput evaluation}
We evaluate the throughput through three deployments: an overloaded deployment with $N_t=4$ and $K=8$, a deployment with $N_t=8$ and $K=8$, and an underloaded deployment with $N_t=16$ and $K=8$. Results are plotted in Fig. \ref{fig:throughput}.

In the first deployment, it is immediately observed that the throughput loss of the SIC receiver with imperfect CSIR increases with the SNR and reached a 29\% loss at SNR=20 dB compared to the SIC receiver with perfect CSIR. In comparison, the MBDL receiver is able to achieve a similar throughput to that of the SIC receiver with perfect CSIR, with an almost negligible loss, throughout the SNR range when 20 training blocks are transmitted using both the minimal and interpolating training mode. This can be explained by the fact that all users receive enough copies of all possible symbol combinations between the common and both private streams during training. Thus, the DNNs are able to learn the full pattern of the interference and noise from the training data. Regarding the interpolating training mode, in 20 training blocks all possible combinations between symbols located in the corners of the employed QAM constellations are still received. This proves to be enough for the DNNs to learn the outer bound of the inter-user interference and achieving similar throughput to using the minimal training mode. As the number of training blocks decreases, the MBDL receiver experiences throughput losses. This is expected as the lack of training data makes it more prone to overfitting to the specific noise and interference pattern of the smaller training set. This is experienced to a greater degree using the interpolating training mode as the error range in the mean symbol calculations increases with fewer training samples. Nevertheless, a large gain compared to the SIC receiver with imperfect CSIR is still observed. In fact, even using the interpolating mode with 5 training blocks offers a 20\% throughput gain compared to the SIC receiver with imperfect CSIR at $\text{SNR}=20$ dB. Thus, this hints at the potential of using the MBDL receiver for robust interference management.

In the second deployment, it is noticed that the MBDL receiver using 20 training blocks with either the minimal or interpolating training mode experiences a slightly greater throughput loss with respect to the SIC receiver with perfect CSIR as the SNR increases compared to the previous overloaded deployment. To understand this effect, it is necessary to take into account that the number of DoFs increases with $N_t$. Thus, when $N_t=8$, the transmitter schedules more private streams compared to relying mainly on the common stream when $N_t=4$. This increase in the total number of private streams makes the interference large and may result in the symbol regions overlapping at the receivers. Thus, the MBDL receiver may incur in errors in the training process and, consequently, during symbol detection in these cases. There is a special observation to be made when using the interpolating pattern with 5 training blocks as for $\text{SNR}<16$ dB, the MBDL receiver achieves lower throughput compared to the SIC receiver with imperfect CSIR. This is due again to the larger interference causing the training symbols to overlap and the overfitting to the small training set, which greatly affects the mean symbol calculations in the interpolation process.

Finally, in the third deployment, the increase to $N_t=16$ allows the transmitter to better manage the interference by spatially separating it with the larger number of DoFs. This allows the MBDL receiver with minimal training with 5 training blocks to outperform the SIC receiver with imperfect CSIR and to achieve similar throughput to the SIC receiver with perfect CSIR when using 20 training blocks. The use of the interpolating training mode also allows the MBDL receiver to considerably outperform the SIC receiver with imperfect CSIR except, again, when using only 5 training blocks due to overfitting at $\text{SNR}<16$ dB.

\begin{figure}[t!]
\begin{minipage}{0.5\linewidth}
\centering
\subfloat[$N_t=4, K=8$]{\label{main:overhead_4_2}\includegraphics[scale=.6]{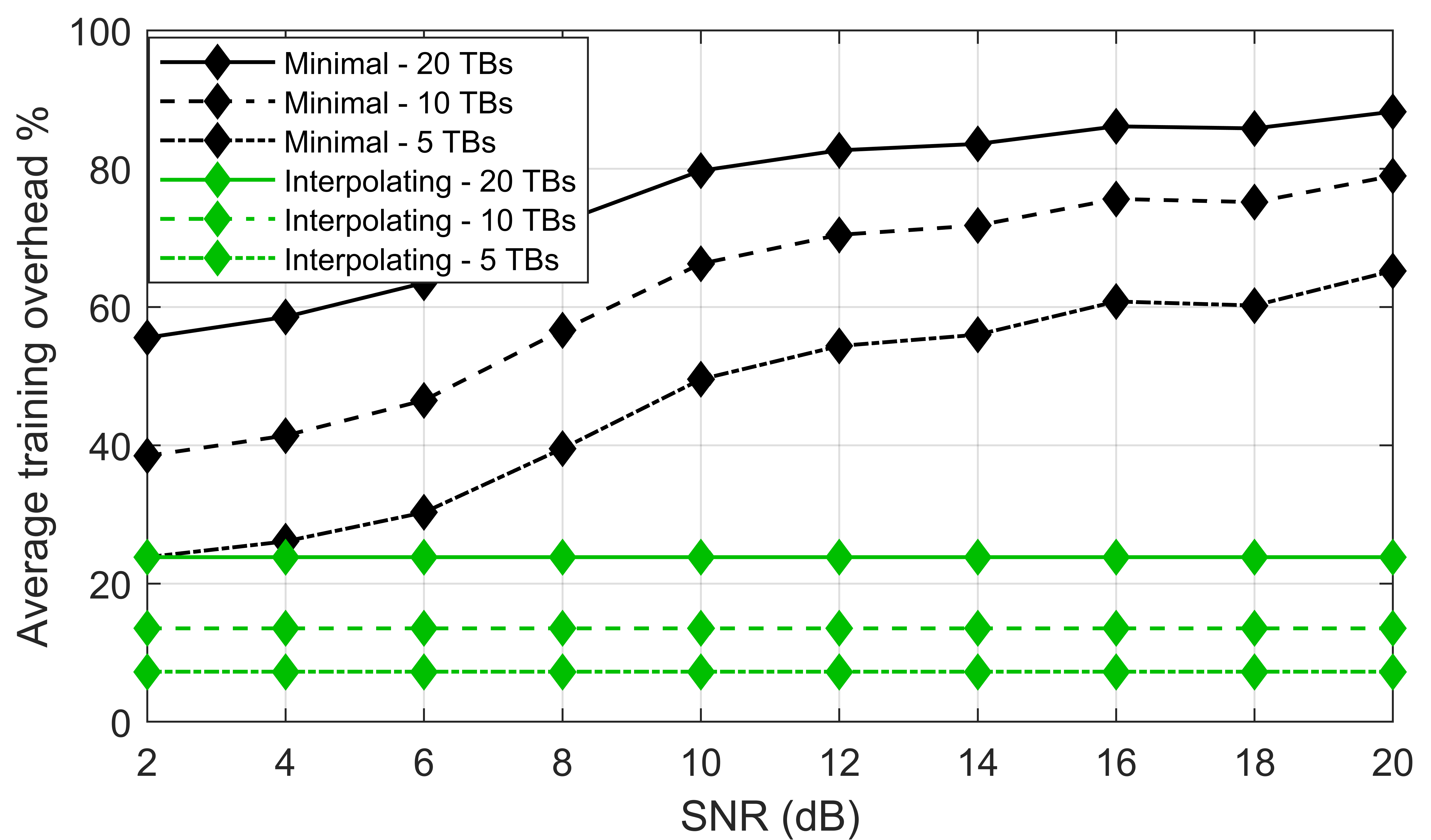}} 
\end{minipage}
\begin{minipage}{0.5\linewidth}
\centering
\subfloat[$N_t=8, K=8$]{\label{main:overhead_8_4}\includegraphics[scale=.6]{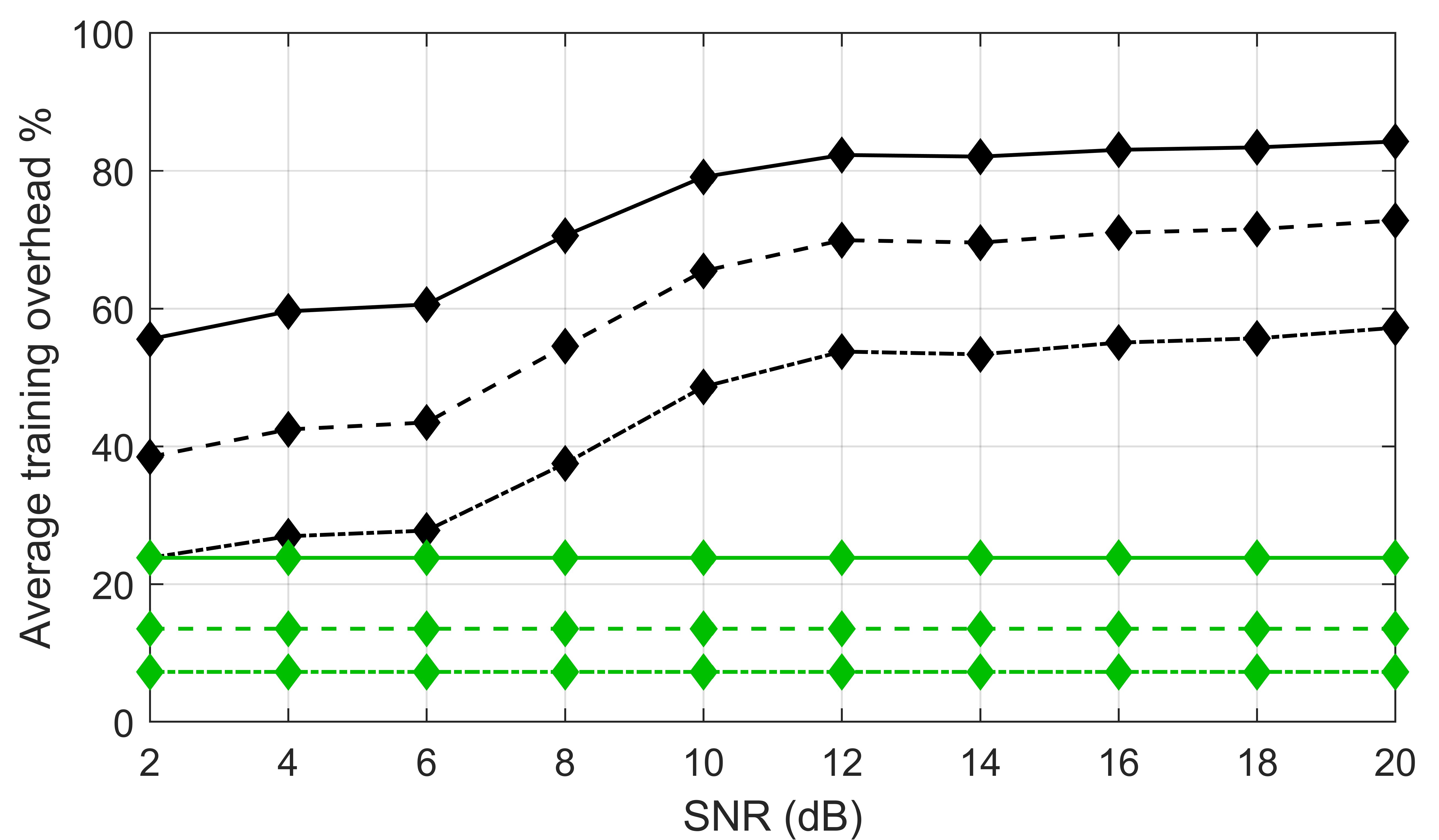}}
\end{minipage}\par\medskip
\begin{minipage}{\linewidth}
\centering
\subfloat[$N_t=16, K=8$]{\label{main:overhead_4_8}\includegraphics[scale=.6]{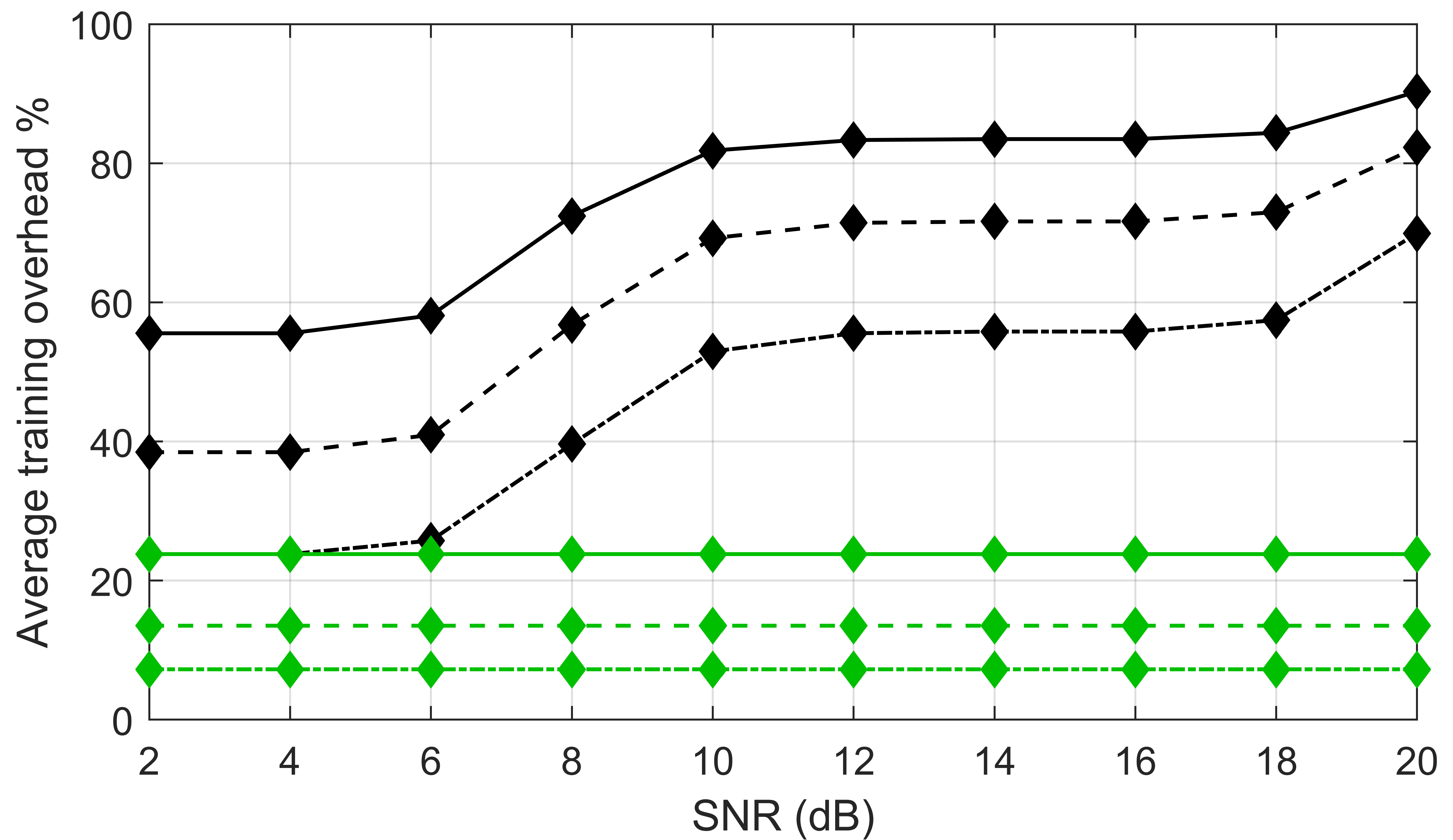}}
\end{minipage}\par\medskip
\caption{Average training overhead vs. SNR.}
\label{fig:overhead_training}
\end{figure}

\subsection{Training overhead evaluation}
The average training overhead curves for the three deployments discussed in the previous subsection are shown in Fig. \ref{fig:overhead_training}. By employing the minimal training mode, the number of transmitted training symbols in each training block increases with the SNR, as the modulation order of the common stream and the user with the highest rate from the precoder optimization process also increase. In turn, employing the interpolating training mode allows the average training overhead to be constant. For 20 training blocks with the interpolating mode, the training overhead is 23.8\%; for 10 training blocks, 15.6\%: and for 5 training blocks, 7.8\%. Thus, employing the interpolation training mode can greatly minimize the training overhead for a practical deployment in linear channels.

\section{Conclusion}
In this work, we introduce a novel MBDL receiver design for RSMA communications to improve the performance over the conventional SIC receiver with imperfect CSIR. Our architecture employs two-DNN banks to simplify the symbol detection task when high order QAM schemes are used, and also to reduce the individual DNN complexity and training overhead. The function of the two compact DNNs in each bank is to take advantage of the symmetry of QAM constellations and classify in which row and column of the constellation the symbol of interest falls in. Furthermore, we propose two training patterns for the MBDL receiver for RSMA communications and we demonstrate that a further training overhead reduction can be achieved compared to transmitting every possible symbol combination between data streams. Finally, through numerical simulations, we demonstrate that the designed MBDL receiver has the potential to perform similarly to the ideal SIC receiver with perfect CSIR in a purely data-driven manner without any prior channel knowledge by generating on-demand non-linear detection boundaries.

Future research directions include a) the expansion of the MBDL receiver to support multi-layer RSMA communications, b) the application of MBDL methods to replace the decoding steps of the common stream and private streams of the MBDL receiver, and c) a real implementation of the MBDL receiver for RSMA communications.

\end{document}